\documentclass[useAMS, graphicx,twocolumn]{mn2e}
\pdfoutput=1

\usepackage{epsf} 
\usepackage{amsmath} 
\usepackage{amsfonts} 
\usepackage{amssymb}
\usepackage{epic}
\usepackage{graphicx}
\usepackage{epsfig}
\usepackage{rotating}
\usepackage{makeidx}
\usepackage{minitoc}
\usepackage{color}
\newcommand{\myfm}[1]{\mbox{$#1$}}
\def\spose#1{\hbox to 0pt{#1\hss}}	% Definition of .le., .ge. symbols
\def\ltabout{\mathrel{\spose{\lower 3pt\hbox{$\mathchar"218$}} % courtesy AJC1.
     \raise 2.0pt\hbox{$\mathchar"13C$}}}
\def\gtabout{\mathrel{\spose{\lower 3pt\hbox{$\mathchar"218$}}
     \raise 2.0pt\hbox{$\mathchar"13E$}}}

\newcommand{\unit}[1]{\ifmmode \:\mbox{\rm #1}\else \mbox{#1}\fi}
\newcommand{\ze}{\ifmmode \mbox{z=0}\else \mbox{$z=0$ }\fi }

\newcommand{\boldv}[1]{\ifmmode \mbox{\boldmath $ #1$} \else 
 \mbox{\boldmath $#1$} \fi}
\newcommand{\half}{\myfm{\frac{1}{2}}}

\def\nb1{{\sf NBODY1} }

\newcommand{\rmd}{\ifmmode \:\mbox{{\rm d}}\else \mbox{ d}\fi }
\newcommand{\rmD}{\ifmmode \:\mbox{{\rm D}}\else \mbox{ D}\fi }

\begin{document}

\title{Black Hole Motion as Catalyst of Orbital Resonances}
\date{2006 December }
% \pubyear{2001} \volume{000} \pagerange{1} % \onecolumn
\author[Boily, Padmanabhan \& Paiement]{C.~M. Boily$^{1,\dagger}$, T. Padmanabhan$^2$ \&  A. Paiement$^{3}$
\\          
$^{1}$Observatoire astronomique de l'Universit\'e de Strasbourg, CNRS, 11 rue de l'universit\'e, F-67000 Strasbourg, France  
\\
$^{2}$IUCAA, Ganeshkhind Post Bag 4, Pune, India 
\\
$^3$E.N.S.P.S. de Strasbourg, Parc d'innovation,  Bd. S\'ebastien Brant
BP 10413, F-67412  Illkirch, France
\\ 
$^{\dagger}$Formerly at: Astronomisches Rechen-Institut, M\"onchhofstrasse 12-14 Heidelberg, 
D-69120 Germany}

\maketitle

\begin{abstract}  
The motion of a black hole about the centre of gravity of its host galaxy induces a strong 
response from the surrounding stellar population. We treat the case of a harmonic potential analytically and show that half of the stars on circular orbits in that potential
 shift to an orbit of lower energy, while the other half receive a positive 
boost and recede to a larger radius. The black hole itself remains on an orbit of 
fixed amplitude and merely acts as a catalyst of evolution of the stellar energy distribution function $f(E)$. 
 We show that this effect is operative out to a radius of $\approx 3 $ to $4$ times the hole's 
influence radius, $R_{bh}$. We use numerical integration to explore more fully the response of a 
stellar distribution to black hole motion. We consider orbits in a logarithmic potential and compare
the response of stars on circular orbits, to the situation of a `warm' and `hot' (isotropic) 
stellar velocity field. While features seen in density maps are now wiped out, 
the kinematic signature 
of black hole motion still imprints the stellar line-of-sight mean velocity to a magnitude $\simeq 18\%$ the local root mean-square velocity dispersion $\sigma$. 
\end{abstract} 
\begin{keywords}{numerical method: N-body;  galaxies, gravitational dynamics}
\end{keywords} 

\section{Introduction} 
Black hole dynamics in galactic nuclei has attracted much attention for many years (e.g., Begelman et al. 
1984; Kormendy \& Richstone 1995; Merritt 2006 for a recent review). 
%The advent of high-resolution spectroscopic cameras in the 1990's, such as the STIS on-board the HST,  
%has led to the identification of a pletora of 
%supermassive black hole (BH) candidates in nearby galaxies (e.g. M87, Kormendy \& Rischstone 1995). 
% Several studies using high-resolution multi-spectral analysis of the kinematics of the central cluster of
% stars around Sgr A$\star$ have confirmed the presence of a massive black hole of $M_{bh} \sim 3\times 
% 10^6 M_\odot$
% at the heart of the Milky Way (MW). 
The influence of a black hole on its surrounding stars is felt first through 
the large
velocity dispersion and rapid orbital motion of the inner-most cluster stars ($\sigma \sim v_{1d} \ltabout 10^3$ 
km/s). This sets a scale $\ltabout GM_{bh}/\sigma^2$ 
($\simeq 0.015-0.019$ pc for the Milky Way, henceforth MW) within which 
 high-angle scattering or  stellar stripping and disruption may take place. For the MW, 
 low-impact parameter star-BH encounters are likely given the 
high density of $\rho \sim 10^7 M_\odot/{\rm pc}^3$ within a radius of 
$\approx 10$ pc (see e.g. Yu \& Tremaine
 2003; O'Leary \& Loeb 2006; see also Freitag et al. 2006 for a   numerical approach to this phenomenon). 
%The recent discovery of high-velocity stars in the MW halo (Brown et al. 2005, 2006) 
%are more easily understood as having originated from such strong interactions with the central massive BH. 
 Star-BH scattering occurring over a relaxation time (Preto et al. 2005 and references therein; Binney \& Tremaine 1987)  leads to the formation of a Bahcall-Wolf stellar cusp of density $\rho_\star \sim r^{-\gamma}$ where $\gamma$ falls in the range 3/2 to 7/4 (Bahcall \& Wolf 1977). 
 Genzel et al. (2003) modeled the kinematics of the inner few parsecs of Sgr A$\star$ 
 with a mass profile  $\rho_\star \sim r^{-1.4}$, suggestive of a strong interplay between the  black hole and 
 the central stellar cusp. More recently, Sch\"odel et al. (2007) presented a double power-law fit to the data, where 
 the power index $\simeq 1.2$ inside a breaking radius $r_{br}$, and $\simeq 7/4$ outside, where $r_{br} \simeq 0.2 $ pc. This is indicative of on-going evolution inside $r_{br}$ not accounted for in the Bahcall-Wolf solution. 
  
Most, if not all, studies of galactic nuclei dynamics assume a fixed black hole (or black hole binary) at the centre of coordinates. 
%On the theoretical side, the orbital evolution of a BH induced by dynamical friction would 
%bring it rapidly to the galactic centre (e.g., Hemsendorf et al. 2002). 
%Radio maps of large-scale outflows as well as high-resolution 
%spectroscopic data, often pointing to compact sources at the photometric centre of the host galaxy, provide indirect support for this ansatz. MW data are more ambiguous. 
Genzel et al. (1997) had set a constraint of $\ltabout 10$ km/s 
for the speed of the black hole relatively to the galactic plane, a  constraint 
later refined to $\ltabout 2$ km/s (Backer \& Sramek 1999; Reid \& Brunthaler 2004). 
%These data bring  support to the traditional picture of a BH frozen at the centre 
%of the Galaxy. Yet new kinematic data puts this paradigm in a new light. 
Stellar dynamics on scales of $\sim $ few pc  surrounding Sgr A$\star$ is complex however, and 
the angular momentum distribution on that scale is a prime example of this complexity (Genzel et al. 2003). 
Reid et al.  (2006) used maser emission maps to compute the mean velocity of 15 SiO 
emitters relatively to Sgr A$\star$. They compute a mean (three-dimensional)
 velocity  of up to 45 km/s, a result obtained from sampling a volume of $\simeq 1$ pc about the centre\footnote{Statistical root-n noise $\sim 25\%$  remains large owing to the small number of sources but is inconsequent to the argument being developed here.}. This raises the possibility that stars within the central  stellar cusp experience significant streaming  motion with respect  to Sgr A$\star$.  
   The breaking radius $r_{br} \sim 0.2$ pc is suggestive of uncertain dynamics 
on that  scale.    Random, `Brownian' black hole motion may result from the expected high-deflection angle encounters (Merritt 2005; Merritt et al. 2007). Here we take another approach, and ask what net effect a black hole set on a regular orbit will have
on the stars. In doing so, we aim to fill an apparent  gap in the modeling of black hole dynamics in dense nuclei, 
by relaxing further the constraint that the hole be held fixed at the centre of coordinates. \newline 

 A rough calculation will help to get some orientation into the problem. 
  Consider  a point mass falling from rest  
from a radius $R_o$ in the background potential of the MW stellar cusp.  
Let the radial mass profile of the cusp $\rho_\star(r) \propto r^{-3/2}$, consistent with  
MW kinematic data. If we define the black hole radius of influence 
$\simeq 1$ pc to be the radius where the integrated mass $M_\star(<r) = $ the black hole mass $ \simeq 3$ to $4 \times 10^6 M_\odot$ (Genzel et al. 2003; Ghez et al. 2005), then $R_o$ may be expressed in terms of the maximum black hole speed in the MW potential as 

\[ \left[ \frac{\max\{v\}}{100\, {\rm km/s}} \right]^{\frac{4}{5}} = \frac{R_o}{1\, {\rm pc}}\, .\]
For a maximum velocity in the range 10 to 40 km/s we find $R_o \simeq 0.3 - 0.5$ pc, 
or the same fraction of its radius of influence\footnote{These figures are robust to details of the stellar
cusp mass profile, so for instance a flat density profile ($\gamma = 0$) would yield $R_o$ 
in the range 0.3 to 0.6 pc.}. We ask what impact this motion might have on the surrounding stars. To proceed further, let us focus on a circular stellar orbit  
 outside $R_o$ in the combined potential of the black hole and an axisymmetric galaxy. 
Fig.~\ref{fig:cartoon} gives a clue to the analysis.  When  the black hole is at 
rest at the centre of coordinates, the star continues on a closed circular orbit of 
radius $r$ and constant velocity $v$. 
We now set the black hole on a  radial path of amplitude $R_o$ down the horizontal x-axis. 
Without loss of generality, let the
angular frequency of the stellar orbit be $\omega^\prime$, and that of the black hole $\omega \ge \omega^\prime$.  The ratio $\omega/\omega^\prime \ge 1$ is otherwise unbounded.
The net force $\bmath{F}$ acting on the star can always be expressed as the sum of
 a radial component $\bmath{F}_r$ and a force parallel to the x-axis which we take to be 
 of the form $F_x \cos(\omega t + \varphi)$; clearly the constant $F_x = 0$ when $R_o =0$. 
 The net mechanical work done on the star by the black hole as the star completes one orbit
  is computed from the integral 
 
 %% eqn Work done by BH in a general way eq:cartoonwork
 %
 \begin{equation}
 \delta W = \int \bmath{F}\cdot\bmath{v} {\rm d}t = \int F_x v \sin(\omega^\prime t) \cos(\omega t + \varphi)\, {\rm d} t
 \label{eq:cartoonwork}
 \end{equation}
where $\varphi$ is the relative phase between  the stellar and black hole orbits. The result of integrating (\ref{eq:cartoonwork}) is best set in terms of the variable $\nu \equiv \omega/\omega^\prime$ as 

\begin{eqnarray}
\frac{\delta W}{r F_x} & = & 
    \frac{1}{\nu^2-1}\left[  \cos(2\pi\nu+\varphi) - \cos(\varphi) \right] 
%%              &   & \ \ + \frac{1}{\nu-1}\left[ \cos(2\pi\nu-\varphi) -\cos(\varphi) \right]
\label{eq:cartoonsolution}
\end{eqnarray}
%
%\begin{equation}
%\frac{2\delta W}{vF_x} = \frac{1}{\nu+1}\left[ \cos(2\pi\nu+\varphi) - \cos(\varphi) \right] 
%+ \frac{1}{\nu-1}\left[ \cos(2\pi\nu-\varphi) -\cos(\varphi) \right]
%\label{eq:cartoonsolution}
%\end{equation}
when $\nu > 1$, and 

\begin{equation}
\frac{\delta W}{r F_x} = \pi \sin(\varphi) 
\end{equation}
when $\nu = 1$. It is a simple exercise to show that this last expression is recovered 
from (\ref{eq:cartoonsolution}) in the limit $\nu\rightarrow 1^+$. Equation (3) 
embodies the essential feature, which is that $\delta W$ changes sign 
when the phase $\varphi$ shifts to $\varphi + \pi$. Thus whenever the stellar phase-space density 
is well sampled and all values of $\varphi : [0,2\pi]$ are realised with equal probability, 
half the stars receive mechanical energy ($\delta W > 0$) and half give off energy ($\delta W < 0$). In other words, stars in the first  quadrant will exchange 
energy with those in the third  quadrant of a Cartesian coordinate system. (Similarly for those in the second and fourth quadrants.) By construction, the black hole neither receives nor loses energy but merely 
acts as a {\it catalyst} for the redistribution of mechanical 
 energy between the stars. Our goal, then, is to explore the consequences of 
this mechanism quantitatively for realistic stellar distribution functions. 

%%fig Cartoon showing the work done through one revolution as function of circular radius r (mass factor M) fig:cartoon
\begin{figure}
\begin{center}
 \begin{picture}(200,220){
%% \put(-50,-135){\epsfig{file=figures/Cartoon.pdf, width=0.6\textwidth}}
 }
 \end{picture}
\end{center}
\caption{Cartoon representing a star on a circular orbit in the combined potential of a black hole and a background galaxy. The black hole motion of amplitude $R_o$ runs parallel to the horizontal x-axis. The net force $\bmath{F}$ acting on the star may be decomposed in a radial component $\bmath{F}_r$ and an x-component.} \label{fig:cartoon} 
\end{figure}

%

%This paper explores how the motion of a massive black hole may affect the velocity distribution function (df) 
%of stars and their spatial distribution. However, instead of focusing on the immediate surroundings of a BH, we 
%look for orbital resonances on a scale of a few times the influence radius of the BH. The aim is to find out 
%what signature would be measurable once the black hole has sunk to the centre and remains fixed there. 
 We begin with an analysis of star-BH orbit coupling in a harmonic (uniform density) galactic potential (\S2). 
While this choice may appear artificial and  an over-simplification, it 
circumscribes all latitude allowed by uncertainties in the spatial distribution of 
stars within the black hole influence radius. Furthermore, the basic mechanics is more
tractable for that case. This is then extended  
 to the case of a logarithmic potential (\S\S3 and 4).  To cover a wider range of parameters, we explore 
with a response code the evolution of individual orbits in the time-dependent  potential. 
 We show that black hole motion shapes up the energy distribution function, as well as  the line-of-sight velocity, which 
 we measure as rms deviations from expected values. The magnitude of these deviations rise 
monotonically with the amplitude of the black hole's orbit, and its mass. Finally, in \S5 we discuss some applications and 
explore possible extensions to our analysis. 

%% Keep for discussion :
%Computer simulations of mergers with seeded black hole suggest that the binary BH forms with an eccentricity 
%that may exceed $e \simeq 0.9$ (Hemsendorf et al. 2002; Makino et al. 2003).  

\section{Circular orbits in a harmonic potential}
 We start with the case of a  star initially on a  circular path in a  background
 harmonic  potential. 
 The star's orbit for that problem is obtained by solving the equations 
of a decoupled oscillator; these read in Cartesian coordinates 

%%eh Harmonic motion equation (vectorial form)  eh:harmonic
\begin{equation} 
\ddot{\boldv{x}} = - \omega^2 \boldv{x}  \label{eq:harmonic}
\end{equation} 
where 

%%eq Definition of the angular frequency omega 
\[ \omega \equiv \sqrt{4\pi\,G\rho/3} \]
is the harmonic angular frequency in an axially-symmetric galaxy of uniform density $\rho$. 
Adding a fixed black hole of mass $M_{bh}$  at the centre of the coordinates preserves the 
symmetry of the force-field:  the motion may still be described 
by (\ref{eq:harmonic}) but with a different angular frequency $\omega^\prime > \omega$. 
Each circular orbit of the harmonic potential maps to a circular 
orbit in this new potential. The aim, then, is to find out what happens once the black hole is 
set in motion, so breaking the symmetry of the force field. 

\subsection{Co-planar, radial BH motion}
We consider a two-dimensional system so all orbits are coplanar. Let the
position of the black hole be denoted $R$ and we take $M_{bh} \gg m_\star $. The black hole 
obeys the same equations (\ref{eq:harmonic}) for any value of $R$ not exceeding the uniform-density 
core of the model potential. (A Bahcall-Wolf cusp would soon develop around the centre once 
the black hole has settled there; the harmonic potential would remain largely unperturbed until that 
happens.) We define the black hole radius of influence $R_{bh}$ such that  

%%eq BH radius of influence eq:rbh
\begin{equation}
M_{bh} \equiv 4\pi \rho R_{bh}^3 / 3 \label{eq:rbh}
\end{equation} 
which we will use again later in a study of the logarithmic potential. 
As a first important case study we set  the black hole on a purely radial path down the x-axis: $\boldv{R} = R\hat{\boldv{x}}$  (from here onwards a hat denotes a unit vector). 
Equation (\ref{eq:cartoonsolution}) gives an expression for the mechanical work 
for the case of a one-dimensional force component. We here develop a second approach based on 
a limited series expansion of the potential. The full  potential 

%%eq Total potential Phi with BH on the x-axis (general expression)
\[ \Phi(r,t) = \frac{\omega^2}{2} r^2 - \frac{GM_{bh}}{\sqrt{(x-R[t])^2 + y^2}} \]
simplifies to 

%%eq Total potential Phi with x-oriented BH (quadrupolar form) eq:quadrupole
\begin{equation} 
\Phi(r,t) = \frac{\omega^2}{2} r^2 - \frac{GM_{bh}}{r} \, \left( 1 - \half \frac{R^2-2xR}{r^2} + O([R/r]^3) \right) \label{eq:quadrupole}
\end{equation}
when we take $r > R$ and truncate a Taylor expansion to leave out all terms of order higher than quadrupolar. In this limit the force acting on the star is easily obtained by differencing (\ref{eq:quadrupole}) 
with respect to $\boldv{r}$. A closed integral through one revolution on the circular stellar path $\boldv{l}$ 
yields the net work done on the star by the time-dependent potential: 

%%eq Work done on star : general path-integral expression eq:work
\begin{equation} 
W = - \int \boldv{\nabla} \Phi {\rm d}\boldv{l} = \int_0^{2\pi} \boldv{\nabla} \Phi\, r{\rm d}\theta\, \hat{\boldv{\theta}}\ . \label{eq:work}
\end{equation} 
The only non-vanishing contribution to the integral comes from the 
non-axially symmetric term of the force field; we find:

%%eq Net work done from the non-axisymmetric term eq:network 
\begin{equation} 
W =  -\int_0^{2\pi} \frac{GM_{bh}}{r^2} \frac{R}{r} r{\rm d}\theta \, (\hat{\boldv{x}}\cdot\hat{\boldv{\theta}}) = \int_0^{2\pi} \frac{GM_{bh}}{r^2} \frac{R}{r} \sin\theta\, r{\rm d}\theta \ .  \label{eq:network}
\end{equation} 
 For circular motion, we may always write $r {\rm d}\theta = v_c {\rm d}t $ with both circular velocity 
$v_c$ and radius $r$ held constant. The ratio $v_c/(\omega r)$ satisfies 

%%eq Ratio of circular velocity to r, omega expressed in terms of M (and definition of M) eq:M
\begin{equation} 
\left(\frac{v_c}{r\omega}\right)^2 = { 1 + \frac{GM_{bh}}{\omega^2r^3}} = 1 + \frac{M_{bh}}{M_g(<r)} \equiv {\cal M} \label{eq:M} \end{equation}
where ${\cal M} \in [1,\infty [$ and 
$M_g$ is the integrated galactic mass inside the orbit {\it of the star}. We have ${\cal M} = 2$ when $r = R_{bh}$ 
by definition, and $\lim_{r\rightarrow\infty} {\cal M} = 1$.    
To progress further requires a specific  form for $R(t)$. Solving (\ref{eq:harmonic}) for a radial orbit 
down the x-axis we find  
 
%%eq Definition of the BH motion equation (harmonic) eq:bhmotion
\begin{equation} 
\bmath{R}(t) = R_o \sin(\omega t + \phi_o ) \hat{x} \label{eq:bhmotion} 
\end{equation} 
where $R_o, \phi_o$ are the amplitude and phase of the black hole orbit. 
The initial conditions are completely specified  
 if we pick the phase of the stellar orbit such that the azimuthal angle 
$\theta = 0$ when $t = 0$. On integrating (\ref{eq:network}), we obtain 

%%eq Final expression of net work W for harmonic BH motion eq:network2
\begin{equation} 
W = - \frac{R_o}{r} \frac{GM_g(<r){\cal M}}{r} \left( \sin[ \frac{2\pi}{\sqrt{\cal M}} + 
               \phi_o] - \sin[\phi_o] \right) \label{eq:network2} \ . 
\end{equation} 
Written in this way we recover a direct proportionality between $W$ and the amplitude $R_o$ (and hence $W = 0 $ 
for zero motion, as expected). $W$ assumes both
positive and negative values according to  the phase $\phi_o$. There can be no correlation 
between $\phi_o$ and the star's orbit, and hence at a given radius $r$ (or, equivalently, binding energy $E$) 
as many stars receive a positive energy increase as those that receive a negative contribution. This impoverishes the occupation level at $r$ (or, $E$) and creates a hollow feature in the stellar distribution 
function.

%%fig Figure showing the work done through one revolution as function of circular radius r (mass factor M) fig:work
\begin{figure*}
\begin{center}
\begin{picture}(200,250){
\put(-185,0){\epsfig{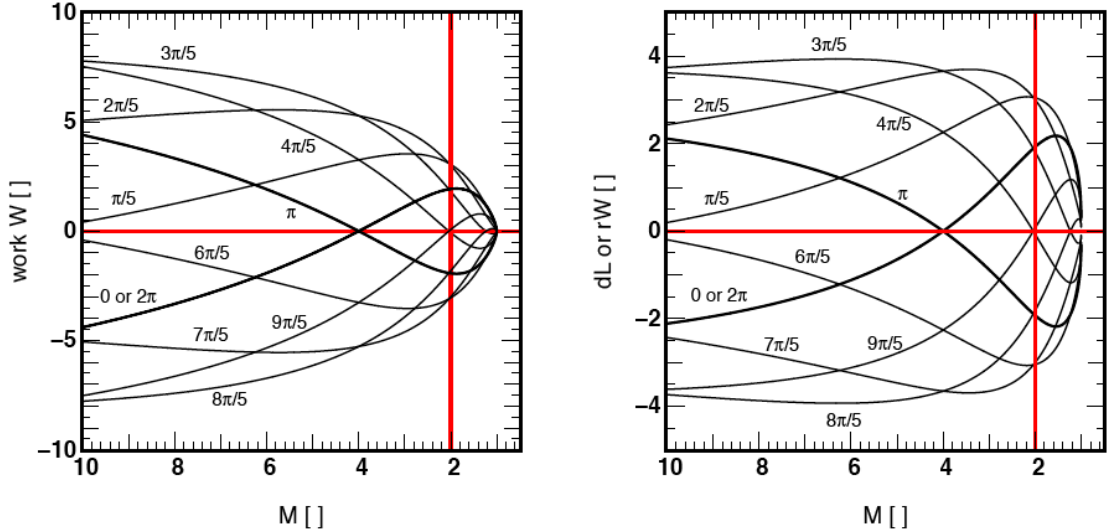}}
%\put(-185,-150){\epsfig{file=figures/WorkvsM.pdf, width=0.7\textwidth}}
%\put(75,-150){\epsfig{file=figures/dLvsM.pdf, width=0.7\textwidth}}
 }\end{picture}
\end{center}
\caption{Left-hand panel: the net work $W$ as function of the dimensionless parameter ${\cal M}$ defined in (\ref{eq:M}). Note that $W$ has been renormalised for better clarity. 
Right-hand panel: the angular momentum ${\rm d}L$ accrued during one revolution as function of ${\cal M}$. 
The curves are labeled with the phase angle 
$\phi_o$. Note that 
the radius $r$ would increase from left to right on the figure. The vertical straight line at ${\cal M} = 2 $ 
corresponds to $r = R_{bh}$, while ${\cal M} = 1$ when $r \rightarrow \infty$. 
 } \label{fig:work} 
\end{figure*}
The work $W$ is a periodic function of $\phi_o : [0, 2\pi]$. In this interval, $W$ vanishes whenever any of the 
following conditions is met:

%%eq Conditions such that W = 0 as function of theta_o  eq:wzero 
\begin{eqnarray} &  &   \left( \frac{2\pi}{\pi - 2\phi_o} \right)^2 \!,\,  \phi_o <\frac{\pi}{2}\ , \nonumber \\
 {\cal M} &=& \left( \frac{2\pi}{3\pi - 2\phi_o} \right)^2 \!, \, \frac{\pi}{2}\le\phi_o\le\frac{3\pi}{2} \label{eq:wzero} \\
& &  \left( \frac{2\pi}{5\pi - 2\phi_o} \right)^2 \!, \, \phi_o >\frac{3\pi}{2} \, . \nonumber
\end{eqnarray}
The full range of values of ${\cal M}$ are covered in the quadrants $-\pi/2<\phi_o<\pi/2$ and 
$\pi/2<\phi_o<3\pi/2$.  As a function of $\phi_o$ we find as many points outside as inside the radius of 
influence ${\cal M} = 2$ (Fig. \ref{fig:work}). 
   For an orbital configuration such that $R_o < r \ll R_{bh}$ a large ensemble of 
solutions to (\ref{eq:wzero}) will cover a wide range of values of ${\cal M}>2$. 
 This will be the case  during  the final stages of a black hole settling at the heart of a galaxy 
since then $R_o \rightarrow 0$. 
It is interesting to 
 investigate whether configurations satisfying (\ref{eq:wzero}) yield islands of 
stability or attractors in phase space. Fig. \ref{fig:work} shows the net work done on a circular stellar orbit
 by the black 
hole as function of ${\cal M}$ for different values of $\phi_o$. To determine whether an orbit will be trapped 
around a point for which $W = 0$, it is sufficient to look at the sign of $W$ in relation to ${\cal M}$. Orbits 
with ${\cal M} < 2 $ lie outside the influence radius $R_{bh}$ at large distances. A gain in binding energy would 
shift the orbit to smaller radii requiring $W < 0$. Similarly, an orbit for which ${\cal M} \gg 1$ would shift to 
higher energy whenever $W > 0$, and so move outwards to larger distance (decreasing {\cal M}). 
Thus curves of $W$ for which ${\rm d}W/{\rm d}{\cal M} > 0 $ when $W = 0$ would trap orbits, otherwise not. 
 We note that the migration induced through $W$ impacts directly on the angular momentum of the star. This is most 
clearly seen from the torque 
$\boldv{\Gamma} = {\rm d}\boldv{L}/{\rm d}t = \boldv{\nabla}\Phi\times\boldv{r} = - GM_{bh}/r^3 R(t) 
\sin\theta \hat{z}$. On integrating over one (stellar) revolution it is easy to show that $|{\rm d}\boldv{L}| 
\propto rW$. At a given radius, the sign of ${\rm d}\boldv{L}$ is a function of $\phi_o$ alone 
(see Fig.\ref{fig:work}[b]). For a well-mixed
systems, half the stars will gain angular momentum and move outwards. Since the background galactic potential 
is taken to be axially symmetric, the distribution of angular momentum at $r$ can then 
be traced back directly to the quality of the black hole orbit (i.e., its amplitude and frequency). 

\begin{figure}
\begin{center}
 \begin{picture}(200,250){
 \put(-60,-150){\epsfig{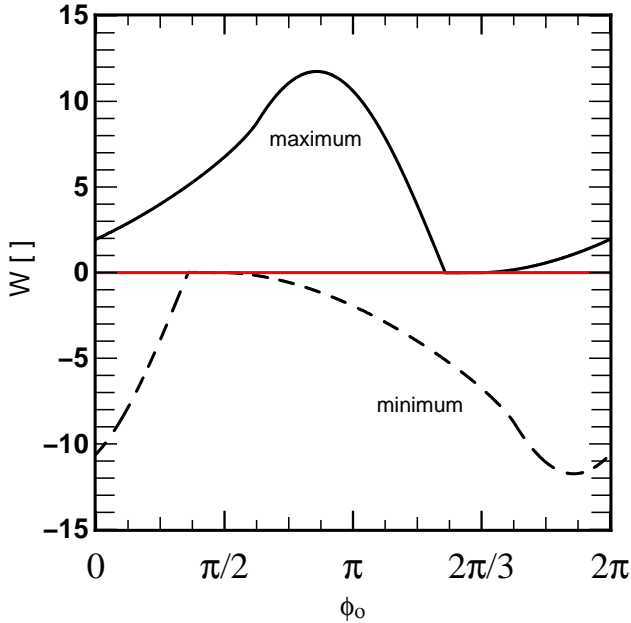}}
%% \put(100,8){\Large $\phi_o$}
 }\end{picture}
\end{center}
\caption{This graphs the maximum (solid) and minimum (dash) value of the work $W$ as a function of the 
phase angle $\phi_o$ (in radians). The two curves coincide when reflected through $W = 0$ and shifted 
by $\delta\phi_o = \pi$ rad.} \label{fig:minmax} 
\end{figure}

It is interesting to map out the minimum and maximum values of $W$ achieved 
as a function of $\phi_o$. 
An extremum  of $W$ occurs when the angle $\phi_o$ satisfies 
$\cos(  2\pi/\sqrt{{\cal M}}  +\phi_o ) = \cos(\phi_o) $ for any constant ${\cal M}$ (or equivalently, $r$). 
Fig. \ref{fig:minmax} graphs the results as a function of the phase angle $\phi_o$. 
The results were obtained by setting a range of ${\cal M}: [1,100]$ for the mass variable. 
 Recall that ${\cal M}\rightarrow \infty $ as $r\rightarrow 0$. The two curves on Fig.~3 coincide if we perform
  a reflection through $W = 0$ together with a shift of $\phi_o$ by $\pi$ radians ($180^{\small o}$). The figure 
shows that when a (positive) maximum is large, the corresponding (negative) minimum is small, and 
{\it vice-versa}. It follows that 
for a finite interval of the phase angle $\phi_o$, for instance, all stars which fall in that interval 
may suffer a net positive (or, negative) intake of energy, irrespective of their orbital radius. 
 Therefore for adequate initial conditions, the energy input from the black hole 
 may leave a signature in the distribution function of the stars 
robust against any bias that might be attributed to domain decomposition (sampling by radius, energy, etc). 
By contrast, stars for which $\max(W) \approx |\min(W)|$ would suffer essentially no effect from 
the black hole. Initial conditions such that $\phi_o \approx \pi/3 $ radians ($\approx 52.7^{\small o}$)  or $\phi_o \approx 4\pi/3 $ ($233^{\small o}$)  are possible examples (cf. Fig.~\ref{fig:minmax}).  

%%fig Figure showing the extrema of W vs angle phi   fig:minmax
By virtue of (\ref{eq:M}) and (\ref{eq:network2}), we find in the limit of large ${\cal M}$ that 
the work $W \propto M^{1/6}\rightarrow\infty$. Hence $W$ diverges as $r\rightarrow 0$. This should not 
come as a surprise since in that limit the orbit is Keplerian around the moving black hole, and hence 
it can not remain close to circular about the centre of coordinates, as we have implied so far. Our 
development will therefore break down when the potential is dominated by the central point source.
With {\cal M = 100}, the largest value considered here,
 the model galaxy contributes 1\% of the dynamical mass only. For the MW galaxy, this would 
translate to a radius around Sgr A$\star$ of $\approx 0.05 $ pc (Genzel et al. 2003). Stellar collisions 
 are predicted to be important on a scale of 0.1 pc (Yu \& Tremaine 2003; O'Leary \& Loeb 2006; Merritt 2006). 
 Therefore, if Milky Way data serve as a test case, we should always set ${\cal M} \ll 100$. 

A final comment concerning the regime ${\cal M} \gg 1$. The definition (\ref{eq:M})
is the ratio of the star's angular frequency to the galactic angular frequency $\omega$. Thus whenever 
${\cal M} \gg 1$, the star revolves rapidly around the black hole. Such trapped orbits remain 
circular to a good approximation when viewed in the reference frame of the black hole. (The eccentricity $e=0$ 
is an adiabatic invariant.) Orbit trapping and resonant relaxation of non-circular Keplerian orbits have 
been discussed by various authors  (e.g.,  Tremaine 1995; Rauch \& Tremaine 1996;  Zhao, Haehnelt \& Rees 2002; Merritt 2006). 
% Clearly  stellar orbits are not circular in general,  
% however, the mechanics of orbit trapping is not at the focus of the present work. These topics have 
% been addressed by various authors: discuss resonances between eccentric stellar orbits trapped around a 
% black hole; and 
% discuss orbital capture in a dense nucleus. 

\subsection{Circular black hole motion}
\subsubsection{Single black hole}
It is straightforward 
to extend the case of radial black hole motion to one where the black hole is on a circular orbit. 
The results of \S2.1 are independent of the sense of rotation of the stellar orbit.  The non-zero  black hole  
angular momentum $\boldv{L}_{bh} = \boldv{R}\times\boldv{p}$ breaks this invariance. The work $W$ will differ 
when stellar and black hole momenta are aligned ($\boldv{L}_{bh} \cdot \boldv{L}_\star > 0$) or anti-aligned 
($\boldv{L}_{bh} \cdot \boldv{L}_\star < 0$), a shown by e.g. Toomre \& Toomre (1972) in their classic study of 
interacting spiral galaxies. 

Consider a black hole on a clockwise two-dimensional circular orbit of radius $R_o$,

%%eq Black hole on clockwise circular motion
\begin{equation}
\boldv{R}(t) = R_o\, ( \sin(\omega t + \phi_o)\, \hat{\boldv{x}} + cos(\omega t + \phi_o  )\, \hat{\boldv{y}}\, ) 
% \equiv \boldv{x}_{bh} + \boldv{y}_{bh}, 
\label{eq:bhcircmotion}
\end{equation}
in a self-evident extension to (\ref{eq:bhmotion}). Repeating the same steps that led to (\ref{eq:quadrupole}) 
will give an extra term for the $y$-component in the potential which is otherwise identical. 
When computing the work (\ref{eq:work}) we may now distinguish between anti- and clockwise stellar orbits
 with the notation ${\rm d}\theta = \pm v_c/r \, {\rm d}t$. The integration yields 

%%eq Final expression of net work W for CIRCULAR BH motion eq:network3
\begin{equation} 
W = - \frac{R_o}{r} \frac{GM_g{\cal M}}{r}(1\mp 1) \left( \sin[ \frac{2\pi}{\sqrt{\cal M}} + 
               \phi_o] - \sin[\phi_o] \right) 
\label{eq:network3}\, .
\end{equation}
This equation shows that aligned orbits double their energy intake from the black hole, while anti-aligned 
ones see a net cancellation. The black hole would, therefore, introduce anisotropy in an initially isotropic
stellar velocity distribution function. This is not unlike the bar amplification process proposed by
 Lynden-Bell (1979): 
the angular frequency $\omega$ set the rotation speed of a constant-magnitude quadrupole, as in a barred galaxy.
However here the perturbation to the axisymmetric galactic potential  is not the two-fold 
symmetric $m = 2$ mode, but the $m=0$ lopsided  mode. 

\subsubsection{Binary black hole}
The case of a binary black hole of constant separation and centered at the origin of coordinates 
is derived from (\ref{eq:network3}) through a 
thought experiment. We imagine that the binding energy of the binary is large and so neglect the 
background galactic potential. We consider the total contribution of the binary to $W$ as a sum
of two single BH's on circular orbits. The phase $\phi_o$ of one black hole (say, the secondary) 
is shifted by $\pi$ radians with respect to that of the primary. On inspection of 
(\ref{eq:network3}) it follows that whenever the product 
$R_o {\cal M}$ is the same for each BH, the total work $W$ must  vanish for any phase angle $\phi_o$. 
This will be the case only when the binary as components of equal mass. 
Since there is no reason to presume identical masses in general,  we would 
find 
\[ W =   - \frac{R_{o,2} - R_{o,1}}{r} \frac{GM_g}{r}(1\mp 1) 
    \left( \sin[ \frac{2\pi}{\sqrt{\cal M}} +  \phi_o] - \sin[\phi_o] \right) \] 
where we have used $ R_{o,1} = R_{o,2}\, M_{bh,2}/M_{bh,1}$. The limit where $ M_{bh,2} \ll M_{bh,1} $ reduces 
to the case of a single BH (the secondary) in orbit in the axi-symmetric potential of the primary  at rest 
at the origin of coordinates. Such a situation might occur when a swarm of intermediate mass black holes 
revolve around a massive hole, presumably the result of repeated coalescence. 
O'Leary \& Loeb (2006) have recently explored the scattering of stars in such a cluster of massive objects. 
 
\section{Case study: the logarithmic potential}
The coupling between black hole motion and orbits in the harmonic potential is 
 indicative of trends that may develop in more realistic potentials. 
Here we recast our problem in the  framework of the logarithmic 
potential, which we write as  

%%eq Definition of the logarithmic potential eq:logarithmic
\begin{equation}
\Phi_g(\boldv{r}) = -\half v_o^2 \ln \left| \frac{x^2+y^2/q^2 + R^2_c}{R_c^2} \right| 
\label{eq:logarithmic}
\end{equation}
with $v_o$ the constant circular velocity at large distances, and $q \le 1$ is a (dimensionless) shape 
parameter. The radius $R_c$ defines a volume inside which the density is nearly constant. Thus 
when $r \ll R_c$ we have once more harmonic motion of angular frequency $\omega = v_o/R_c$. 
If we let $ q = 1 $ and define $ u \equiv r/R_c$, the volume density $\rho$ reads 

%%eq Density of logarithmic potential eq:logrho 
\begin{equation}
4\pi G\rho(u) = \nabla^2\Phi_g = \frac{v_o^2}{R_c^2} \frac{3 + u^2}{(1 + u^2)^2} 
\label{eq:logrho}
\end{equation}
and the integrated mass $M_g(u)$

%%eq Integrated mass of the log potential eq:logmass 
\begin{equation}
M_g(<u) = \frac{v_o^2R_c}{G} \frac{u^3}{u^2+1} \, .
\label{eq:logmass}
\end{equation}
The mass $M_g(u\gg 1) \propto u$ diverges at large distances, however this is not a
serious flaw since we will consider only the region where $u \sim 1$. The mass 
$M_g(u=1) = v_o^2R_c/2G$ fixes a scale against which to compare the black hole mass $M_{bh}$. 
Since the black hole will orbit within the harmonic core, we set 

%%eq Define mbh tilde (ratio of Mbh to M(core) of the galaxy eq:mbhtilde 
\begin{equation}
M_{bh} \equiv \tilde{m}_{bh} \frac{v_o^2R_c}{2G} 
\label{eq:mbhtilde}
\end{equation}
with $0 < \tilde{m}_{bh} \le 1$, and 

\[ {\cal M}(u) = 1 + \tilde{m}_{bh} \frac{1+u^2}{u^3} \] 
bears the same meaning as before. 
The core radius offers a reference length to the problem. 
The position and velocity of the black hole at any time 
 follow from (\ref{eq:bhmotion}), where the amplitude is set by fixing 
 the dimensionless number $u_o = R_o / R_c$ and the angular frequency $\omega$ is
 
 \[ \omega = \frac{v_o}{R_c} \ . \]
Our goal is to quantify the 
 time-evolution of a large number of orbits in the combined logarithmic and 
 black hole potentials. If we pick parameters such that 
 
 \[ m_\star \ll M_{bh} < M_g(\max\{u\}) \]
 then we may neglect the collective feedback of the stars on the black hole and 
 galactic potential and study only the response of individual orbits evolving in the 
 time-dependent total potential. This approach will remain valid so long as the 
 response of the stars are relatively modest. The time-evolution of orbits was done 
 numerically using a standard integration scheme, which we describe below. 

\subsection{Equations and numerics}
The energy $E_J$ of a star is computed from 

%%eq Energy of the orbit  eq:jacobi
\begin{equation}
E_J = \half v^2 + \Phi_g(\bmath{r}) + \Phi_{bh}(\bmath{r}, t) 
\label{eq:jacobi}
\end{equation}
where 

%%eq Black hole potential (general form) eq:bhpotential
\begin{equation}
\Phi_{bh}(\bmath{r},t) \equiv - \frac{GM_{bh}}{|| \bmath{r}-\bmath{R}||} 
\label{eq:bhpotential}
\end{equation}
is an explicit function of time through (\ref{eq:bhmotion}); from here onward we will write $(x_{bh},y_{bh})$ for 
the coordinates at time $t$ of the black hole in the $xy$-plane. The time-derivative 

\[ \dot{E}_J = \partial_t \Phi_{bh}(\bmath{r},t) \neq 0 \]
and hence energy is not a conserved quantity. We use this fact to define a set of six 
first-order differential equations 

%%eq Set of equations to solve numerically eq:equations
\begin{equation}
\frac{\rm d}{{\rm d}t} {(\bmath{r}, E_J, \bmath{v}, \dot{E}_J)} = (\bmath{v}, \dot{E}_J, -\bmath{\nabla}\Phi, F_J)
\label{eq:equations}
\end{equation}
where the time-derivatives are computed in the usual way, and we find for a black hole orbit 
confined to the harmonic core 

%%eq Definition of F_J  eq:fj 
\begin{eqnarray}
F_J  & = & \nonumber \\ 
 &3&\!\!\!\!\!\!\!\frac{GM_{bh}}{||\bmath{r}-\bmath{R}||^{5}}\,
      \left( [x_{bh}-x]\,[\dot{\widehat{x_{bh}-x}}] +   [y_{bh}-y]\,[\dot{\widehat{y_{bh}-y}}] \right) \nonumber \\ 
  & +&\!\!\!\!\! \frac{GM_{bh}}{||\bmath{r}-\bmath{R}||^{3}} \, 
  \left( \dot{x}^2_{bh} +  \dot{y}^2_{bh} - \dot{x}\dot{x}_{bh} - \dot{y}\dot{y}_{bh} \right) \nonumber \\
  & -&\!\!\!\!\!\omega^2\,  \frac{GM_{bh}}{||\bmath{r}-\bmath{R}||^{3}} \,  
  \left( [x_{bh}-x] x_{bh} + [y_{bh}-y] y_{bh}\right) \, .
\label{eq:fj} 
\end{eqnarray}
The expression for $F_J$ admits a simplification
 when the black hole is set on a radial orbit (\ref{eq:bhmotion}) but note 
that (\ref{eq:fj}) is not invariant to a  swap of  $x$ for $y$ and {\it vice versa} when $y_{bh} = 0$. 

\subsubsection{Compact kernel}
The potential (\ref{eq:bhpotential}) is singular when $\bmath{r} = \bmath{R}$ which introduces large errors 
in the integration. To alleviate this we redefine (\ref{eq:bhpotential}) using a compact kernel, 
effectively  smoothing over the singularity. Let $\varepsilon_{bh}$ be a constant length and write 

%%eq Definition of epsilon as compact kernel length eq:epsilon
\begin{equation}
\Phi_{bh}(\bmath{\cal R},t) = - \frac{GM_{bh} }{ \varepsilon_{bh} } \hat{\Phi}(\bmath{\cal R},t),  
\label{eq:epsilon}
\end{equation}
whenever ${\cal R} \le 1$ where ${\cal R} = ||\bmath{r}-\bmath{R}||/\varepsilon_{bh}$. 
We pick a compact kernel which minimises force errors at ${\cal R } = 1$  (Dehnen 2003) and define 

%%eq Definition of compact kernel eq:kernel 
\begin{equation} 
\hat{\Phi}({\cal R}) = 1 + \half\,( 1 - {\cal R}^2 ) + \frac{3}{8}\,(1 - {\cal R}^2)^2  \ . 
\label{eq:kernel} 
\end{equation}
This last equation fails when ${\cal R} > 1$, however this is of no concern since the gradient is continuous 
at ${\cal R} = 1$ and matches exactly the one derived from (\ref{eq:bhpotential}) at that radius. Integration 
of equations (\ref{eq:equations}) with respectively (\ref{eq:bhpotential}), or (\ref{eq:epsilon}) and (\ref{eq:kernel}), when ${\cal R} > 1$ or $\le 1$,  poses no particular difficulty, 
 though Eq.~\ref{eq:fj} takes another form inside ${\cal R} < 1$ (see below). 

\subsubsection{Choice of units \& integrator}
For convenience we have chosen scales for the background potential such that $G = v_o = M = 1$. Borrowing from the case of the 
MW black hole, we set a kernel length $\varepsilon_{bh} = 2\times 10^{-2}$ which will wipe out all high-deflection angle collisions, i.e., those due to orbits with little angular momentum. 

We have used an explicit fourth-order time-adaptative Bulirsch-Stoer integrator taken from Press et al. (2002) 
for solving (\ref{eq:equations}).
 We have performed a series of tests with a static potential by setting e.g. $\tilde{m}_{bh} = 1$ and $u_o = R_o = 0$ in (\ref{eq:bhmotion}). With these parameters $\dot{E}_J = 0$, and we checked that a precision of 
 $1:10^{14}$ is 
 maintained for a runtime of 200 units. In particular we validated Eq.~\ref{eq:kernel} by integrating radial stellar 
 orbits running through the black hole, both along the x- and y-axis. We also integrated a circular orbit  at 
 the edge of the kernel, or ${\cal R } = 1$, and found no indication of a drift in energy or any kind of random 
 fluctuations. 
 
 The situation is less glorious when integrating with finite black hole motion. To see why, let us write 
 down $F_J$ in (\ref{eq:equations}) with (\ref{eq:bhmotion}) and ${\cal R } < 1$. Some straight forward 
 algebra yields 
 
 %%eq Equation of FJ inside kernel eq:fjkernel 
 \begin{eqnarray}
 F_J &=& 3 \frac{GM_{bh}}{\varepsilon^5_{bh}} \, \left( x-x_{bh}\right)\, (\dot{x}-\dot{x}_{bh}) \nonumber \\
& & - \frac{GM_{bh}}{\varepsilon^3_{bh}} \left( \frac{5}{2} - \frac{3}{2} {\cal R}^2\right) \frac{\rm d}{{\rm d}t} (x-x_{bh})\dot{x}_{bh} \, .
 \label{eq:fjkernel}
 \end{eqnarray}
 Since the relative distance between the star and the black hole is $< \varepsilon$, we find that $F_J < O(1/\varepsilon^5) $ which gives $O(10^9)$ for the  parameters chosen. As such this would not be problematic, 
 however the time-steps required to maintain an accuracy of $1:10^{14}$ for a typical integration time 
 become tiny, and the computer run-time, prohibitive. We found a practical solution to this problem, 
 by imposing
 that the quantity $\varepsilon^3 F_J$ be integrated to a precision of $2:10^{12}$, so that $F_J$ is known to 
 six significant digits ($\varepsilon \sim 10^{-2}$). Whenever this condition was not met, we only 
 included the orbit in the  analysis {\it up to} that point in time, after which it was ignored. 
 This situation occurred relatively  seldom, and affected 
 some $3 - 4\%$ of cases at most. This was so, for example, when the initial configuration either started out with many stars on near-radial orbits and small-amplitude black hole motion; or, when the black hole was allowed to flirt with a large number of stars by covering a distance comparable to or exceeding  its radius of influence. 
 
\subsection{Initial conditions} 
We limit our exploration to the case of coplanar motion. The volume density (\ref{eq:logrho}) 
stretches to infinity and yields a divergent integrated mass. Since we are only interested in the 
central-most volume, we decided to keep only stars that remain inside a given radius. (A selection by 
energy $E_J$ would be equivalent.) Our model calculation of \S1 had shown that the response of the 
star to black hole motion is a strong function of the ratio of their orbital periods. The orbital period of the 
star is $\propto 1/\sqrt{G\overline{\rho}} $, where $\overline{\rho}$ is the mean volume density inside the 
semi-major axis of the orbit. These two observations combined  suggests that we only include orbits out 
to where the density varies most rapidly. We chose a truncation radius such that $u_t \le e$, close to the 
value 2.80(7)  at which ${\rm d}\ln\rho/{\rm d}\ln r = -2.101 .. $ reaches a minimum. \\

We now specialise to the case of circular orbits in the axisymmetric  potential $\Phi_g$ by setting $q = 1$ 
in (\ref{eq:logarithmic}). Positions are attributed by Monte Carlo method using the density as probability 
distribution. The square circular velocity at each radius is 

%%eq Circular velocity in the log potential eq:logvc 
\begin{equation}
\left(\frac{v_c}{v_o}\right)^2 = \frac{u^2}{u^2+1} + \frac{\tilde{m}_{bh}}{2u} 
\label{eq:logvc}
\end{equation}
and the sense of motion chosen randomly so that the total angular momentum of the stars is zero to 
within root-n noise. The energy may be written as 

%%eq Energy for log and circular motion eq:energylogvc
\begin{equation}
\frac{2E_J}{v_o^2} = \ln|u^2+1| + \frac{u^2}{u^2+1} - \frac{\tilde{m}_{bh}}{2u} \, .
\label{eq:energylogvc}
\end{equation}
This expression is easily differentiated to yield an analytical form for $E^\prime_J = {\rm d}E_J/{\rm d}u$ which 
is the density of states of stars of energy $E_J$ at $u$. 
Since we are only interested in coplanar orbits, the mass element drawn from (\ref{eq:logrho}) 
is 
\[ \delta M = 2\pi R_c^3 \rho(u) u {\rm d}u \equiv f(u) {\rmd }u = f(u) \frac{{\rm d}E_J}{E^\prime_J} \equiv f(E_J) {\rm d}E_J \] 
where the energy distribution function $f(E_J)$ is known in parametric form, 

%%eq Energy distribution function for circular motion eq:dfelog
\begin{equation}
f(E_J) \equiv \frac{R_c v_o^2}{G} \frac{ u^3\, (3+u^2)}{\tilde{m}_{bh} (u^2+1)^2 + 4u^3\,(u^2+2)} \, .
\label{eq:dfelog}
\end{equation}
This equation shows that when  $\tilde{m}_{bh} = 0$ (no black hole) we find $f(E_J) \rightarrow $ constant in the 
limit $u\rightarrow 0$; and $f(E_J) \propto u^3$ in the same limit when $\tilde{m}_{bh} \ne 0$. Thus the bulk 
of the stars avoid the central black hole. Equation (\ref{eq:dfelog}) will be  helpful when assessing 
the noise level of the response of the stars to black hole motion. 
 
\section{Results} 
\begin{figure}
\begin{center}
 \begin{picture}(200,250){
\put(-60,-150){\epsfig{file=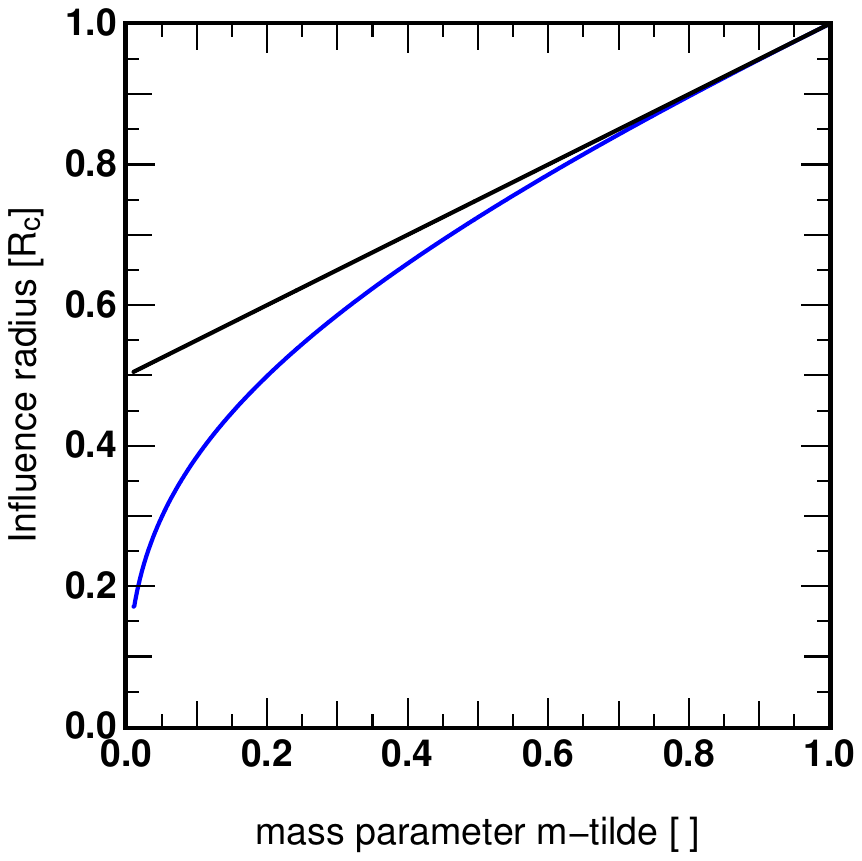, width=0.7\textwidth}}
} \end{picture}
\end{center}
\caption{Influence radius in units of $R_c$ 
as a function of the mass parameter $\tilde{m}_{bh}$ defined in (\ref{eq:mbhtilde}). 
In the limit $\tilde{m}_{bh}\rightarrow 0$ the radius $\propto \tilde{m}_{bh}^{1/3}$ rises rapidly. In the neighbourhood 
of $\tilde{m}_{bh} = 1$, the radius $\propto \tilde{m}_{bh}$; the straight line is the curve $\tilde{m}_{bh}/2 + \half$. 
 } \label{fig:influenceradius} 
\end{figure}

We have until now fixed the gravitational constant $G = 1$ and velocity scale of the 
galactic potential $v_o = 1$. The initial conditions require further that we fix the black hole mass parameter $\tilde{m}_{bh}$ in (\ref{eq:mbhtilde}), black hole's amplitude of motion $u_o$, and the core radius, $R_c$. The total  
mass (\ref{eq:logmass}) will integrate to 1 up to $u_t \simeq e$ if we set $R_c \simeq 0.46$. This fixes all scales in the 
problem, and we note that $\approx$ half the stars lie inside $R_c$, and an equal fraction outside. 
The black hole's radius of influence is obtained in terms of $\tilde{m}_{bh}$ 
from equating (\ref{eq:logmass}) to (\ref{eq:mbhtilde}). The result is shown on Fig.~\ref{fig:influenceradius}. 
 We use this relation to set  a more stringent 
 constraint on the motion of the black hole by imposing that it orbits always within its radius of influence, 
  an improvement on our initial ansatz that  $u_o < 1$, since (\ref{eq:bhmotion}) is a better solution 
  to the black hole's orbit closer to the origin. 
 
%%fig Figure showing the influence radius as function of mtilde  fig:influenceradius

%%tab List of initial conditions tab:initialconditions
\begin{table} 
\caption{Initial conditions of the numerical orbit integrations.  
The influence radius $r_{bh}$ defined in (\ref{eq:rbh}) is given in computational units. The potential (\ref{eq:logarithmic})
is defined in units such that $G = v_o = 1$. The core radius $R_c = 0.459$,  and the truncation radius $r_t \simeq 2.53$   gives an integrated mass (\ref{eq:logmass}) = 1. 
\label{tab:initialconditions}  }

\begin{center} 
\begin{tabular}{lcccc}  
\multicolumn{5}{c}{Circular stellar orbits (`cold')} \\ 
Name  & $\tilde{m}_{bh}$ & $u_o $ &  $r_{bh}$ & Comment \\ 
                            &    &   $[R_o/R_c]$       &           &          \\
C1        & 0.30                       &        0.00    &       0.258    &              Static \\
C2        & 0.30                       &        0.07     &      0.258    &            \\
C3        & 0.30                       &        0.15    &    0.258       &      Reference case       \\
C4        & 0.30                       &        0.21     &  0.258       &               \\
C5        & 0.30                       &        0.30      & 0.258         &                 \\
\\
C2s     & 0.15                       &        0.15      &   0.197      &            Shadows C2     \\
C3s     & 0.15                       &        0.30      &    0.197     &            Shadows C3    \\
\\
\multicolumn{5}{c}{`Warm'  or `Hot' runs } \\ 
Name  &  $\tilde{m}_{bh}$ & $u_o$ &  $r_{bh} $ & Comment \\ 
                             &     &    $R_o/R_c$       &    &                          \\
W1       & 0.30                       &         0.15      & 0.258         &            \\
W1c      & 0.30                       &         0.15   & 0.258            &             circular BH orbit \\
H1        & 0.30                       &         0.15    & 0.258           &            \\
\\
W2       &0.30                       &         0.21     & 0.258          &            \\ 
\end{tabular}
\end{center} 
\end{table}

\subsection{Reference case, $\tilde{m}_{bh} = 0.3$} 
We set up a reference case inspired by MW data which will guide us through our exploration of parameter space. 
Since the MW black hole lies close to the galactic centre at a velocity $\simeq 2-4 $ km/s
this suggests that we focus on cases where the black hole remains well inside the central core. We pick a black hole 
mass equal to  30\% of the core mass, $\tilde{m}_{bh} = 0.3$, and set an upper limit of $u_o = 0.66$ on its amplitude of
motion, when it would exceeds slightly its influence radius of $\simeq 0.57$ (Fig.~\ref{fig:influenceradius}). 

%%fig Figure showing Poincare sections  fig:poincare
\begin{figure*}
\begin{center}
 \begin{picture}(200,550){
 \put(-75,60){\epsfig{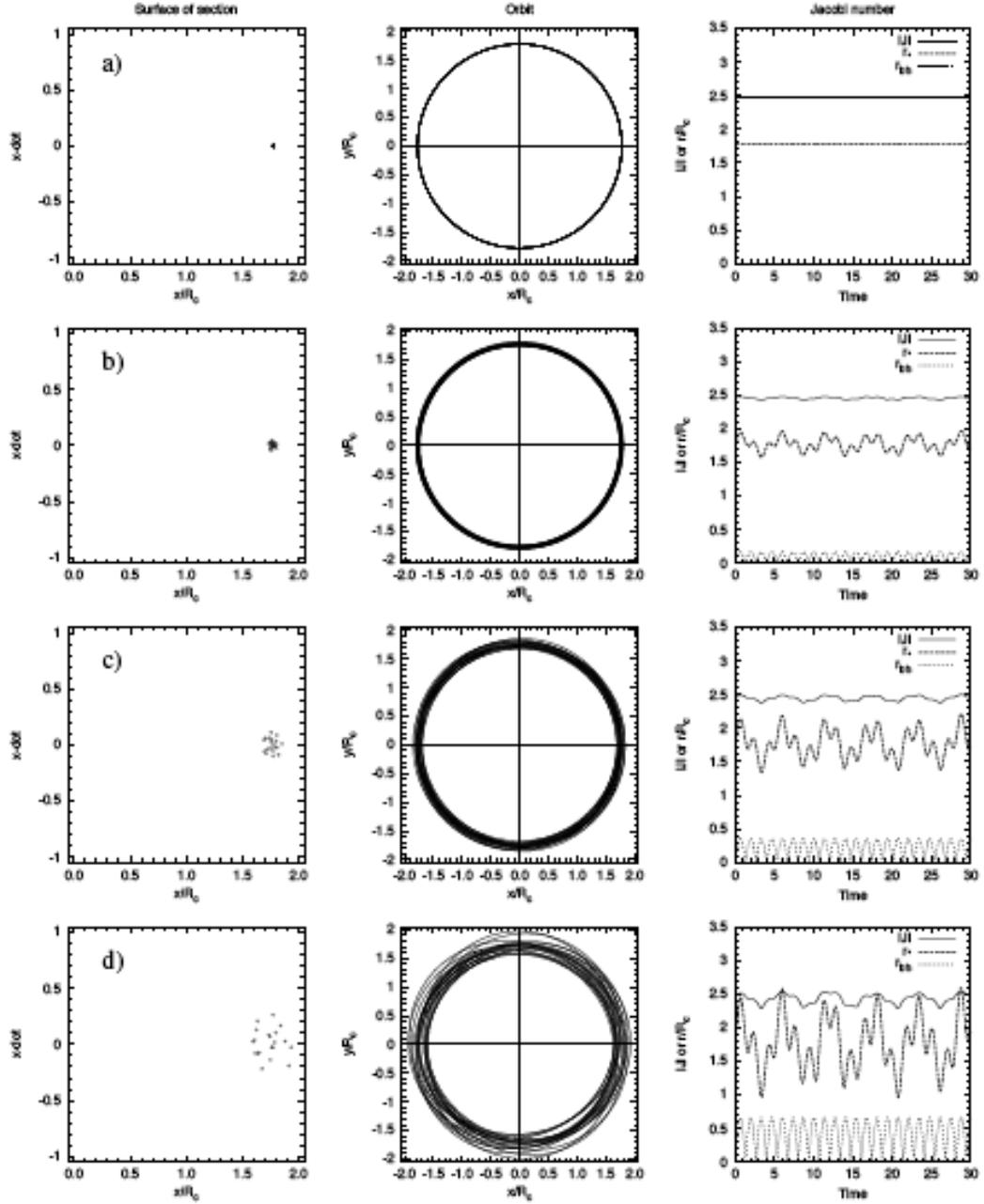}}
% 
% \put(-25,330){\epsfig{file=figures/Section0b.pdf, width=0.5\textwidth}}
% \put(125,330){\epsfig{file=figures/Section0c.pdf, width=0.5\textwidth}}
%	\put(-175,195){\epsfig{file=figures/Section1a.pdf, width=0.5\textwidth}}
% 	\put(-25,195){\epsfig{file=figures/Section1b.pdf, width=0.5\textwidth}}
% 	\put(125,195){\epsfig{file=figures/Section1c.pdf, width=0.5\textwidth}}
% \put(-175,60){\epsfig{file=figures/Section2a.pdf, width=0.5\textwidth}}
% \put(-25,60){\epsfig{file=figures/Section2b.pdf, width=0.5\textwidth}}
% \put(125,60){\epsfig{file=figures/Section2c.pdf, width=0.5\textwidth}}
%	 \put(-175,-75){\epsfig{file=figures/Section3a.pdf, width=0.5\textwidth}}
%	 \put(-25,-75){\epsfig{file=figures/Section3b.pdf, width=0.5\textwidth}}
%	 \put(125,-75){\epsfig{file=figures/Section3c.pdf, width=0.5\textwidth}} 
 }\end{picture}
\end{center}
\caption{Poincar\'e section ($\dot{x},x$) at $y = 0 $ (left panels), orbit (middle) and radii and energy (right) for 
a single stellar orbit in the logarithmic potential to which we added an $\tilde{m}_{bh} = 0.3$ black hole. 
The panels to the right display the binding energy of the star (solid line) along with the distance $r_\star$ of the star to the black hole (dash). Modulations in energy match one to one variations in $r_\star$.  The black hole orbit $r_{bh}$ (dots) is 
also displayed for comparison. The four rows show the orbit for different values of $u_o$: a) 0.0, b) 0.15, c) 0.33 and d) 0.66. 
 } \label{fig:poincare} 
\end{figure*}

To see how orbits respond as $u_o$ is increased from zero, we draw on figure \ref{fig:poincare}  Poincar\'e sections of a single orbit for four values of $u_o: 0, 0.15, 0.33 $ and $0.66$. The case of a fixed black hole is shown on Fig.~\ref{fig:poincare}(a), when the star describes a circular orbit of radius $r = 1.78$ which is $\gtabout 3$ times the influence radius 
of the black hole. The orbital period $ = 2\pi r/v_c = 11.18$ and the integration was for a total of 200 time units (18 revolutions). The middle- and right panels show the orbit and the star's energy (solid curve) and distance to the black hole (labeled $r_\star$, dashed curve), respectively. Fig.\ref{fig:poincare}(c) illustrates the situation 
when the black hole motion has amplitude $u_o = 0.33 \simeq 0.58 \times $ its influence radius.  The 
star's orbital radius now 
varies from a minimum of approximately 1.65 and up to 1.85, a gap of $\approx 10\%$ compared to the circular orbit; 
the same conclusion applies to the 
cycles seen in binding energy (right panel, Fig.~\ref{fig:poincare}[c]). We note that the modulations in $E$ match one to one 
the profile of $r_\star$, which is an indication that a strong coupling is still operative even for stars orbiting well beyond
the black hole's radius of influence. The scatter seen in the Poincar\'e sections of both Fig.~\ref{fig:poincare}(c) and (d) 
confirms this view, and suggests that the original circular orbit becomes mildly chaotic (no loop or resonant structure). \\
%We will not pursue this aspect further in this contribution. \\

 %%fig Figure showing the energy distribution function for four values of N  fig:noiselevel
\begin{figure*}
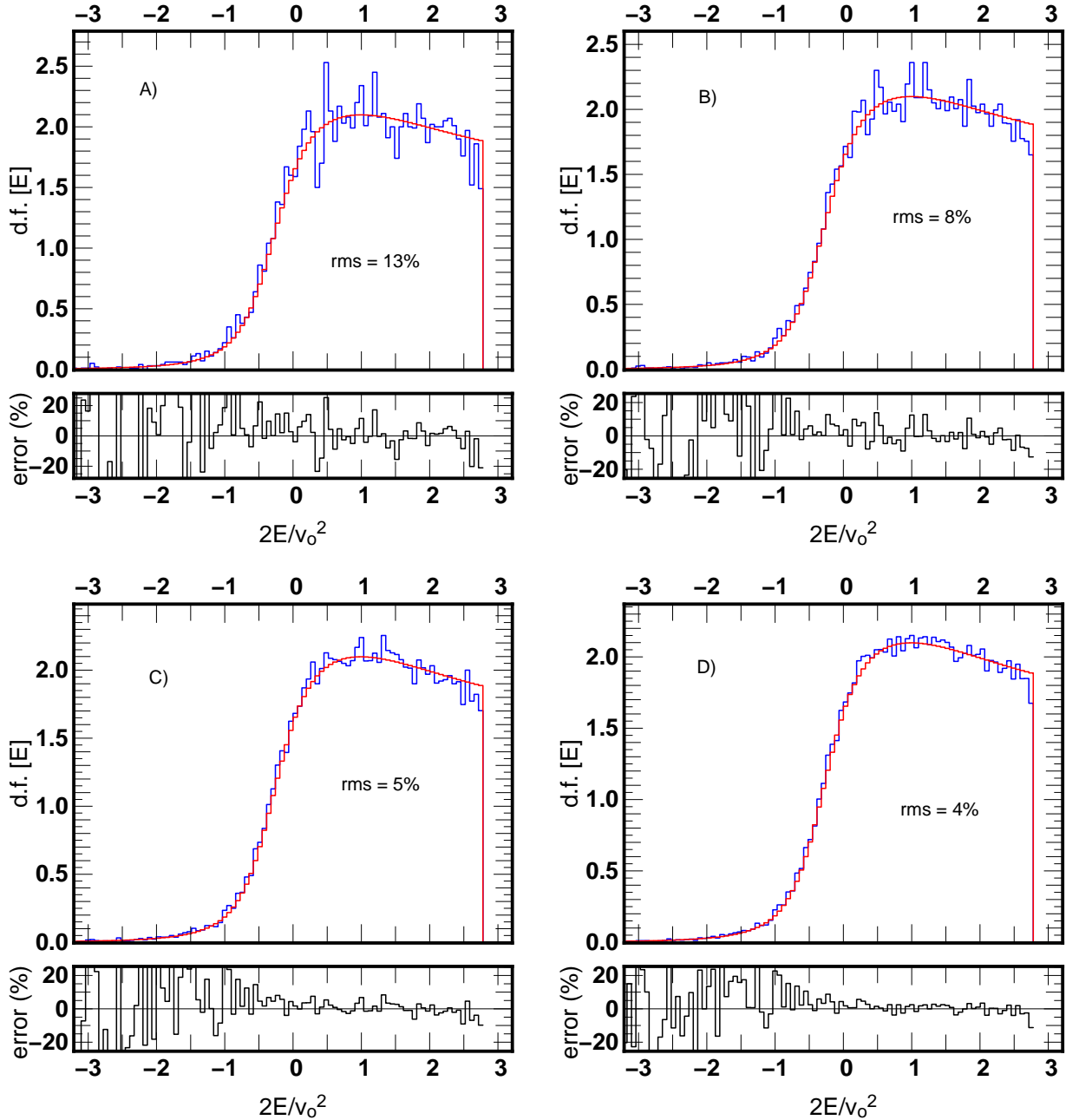

\begin{center}
 \begin{picture}(200,500){
 \put(-200,100){\epsfig{file=figures/dfE_a.pdf, width=0.7\textwidth}}
 \put(40,100){\epsfig{file=figures/dfE_b.pdf, width=0.7\textwidth}}
 \put(-200,-150){\epsfig{file=figures/dfE_c.pdf, width=0.7\textwidth}}
	\put(40,-150){\epsfig{file=figures/dfE_d.pdf, width=0.7\textwidth}}
 }\end{picture}
\end{center}
\caption{The energy distribution function (d.f.) for four discrete realisations with (from [a] to [d]): $N = 10,000, 20,000, 40,000$ and  $80,000$ orbits. In each case the span in energy was divided in 200 bins of equal size. 
The analytic curve (\ref{eq:dfelog}) is displayed as the more regular histogram on each panel, and the root-mean square 
differences given. The rectangular frames gives the relative differences (in \%) for each bin.} \label{fig:noiselevel} 
\end{figure*}

The response of individual orbits to black hole motion would leave a measurable trace only if their signal rises above 
the background noise of other orbits that may be otherwise affected. We sought out a relation between an unperturbed 
and time-independent distribution function and the noise level of a discrete realisation of that function with $N$ bodies.
The reference distribution function is given by (\ref{eq:dfelog}) which we discretised using 100 equal-size  bins of width $\Delta E \simeq 0.07$. This curve is plotted as a histogram 
 on Fig.~\ref{fig:noiselevel}. We then measured the root-mean square differences with the analytic distribution function 
 by drawing different numbers of orbits in the range 10,000 to 100,000. The results are shown for four values of $N$ on 
 Fig.\ref{fig:noiselevel}(a-d). We find the rms differences drop to $\sim 4-5\%$ already for $N \gtabout 40,000$, comparable 
 to Poisson noise (roughly $1/\sqrt{400} \simeq 5\%$ fluctuations per bin). For completeness, we also plot the relative differences in percentage 
 at each bin of energy on the rectangular frames below each panel. 
 This relative energy error is dominated by low-number statistics and becomes very large when $E \ltabout -1$, which, for 
 our choice of parameters, corresponds to a radius $u \simeq 0.074$ enclosing $\simeq 0.02\%$ of the total mass (expected Poisson noise of $\approx 20\%$). The error 
 made in dropping orbits below that level of energy from our analysis is of the same order. Furthermore, stars that are that 
 close or closer to the origin would be trapped by the black hole gravity and remain on high-velocity 
 Keplerian orbits around it. This would remain true  even if the black hole were set in motion at a comparatively small velocity. Such orbits are not  the focus of this work. 

%%fig  Energy distribution function of an m= 0.3 system @ t = 20 time units of evolution.  fig:signal
%%fig   fig:timeaverage
\begin{figure*}
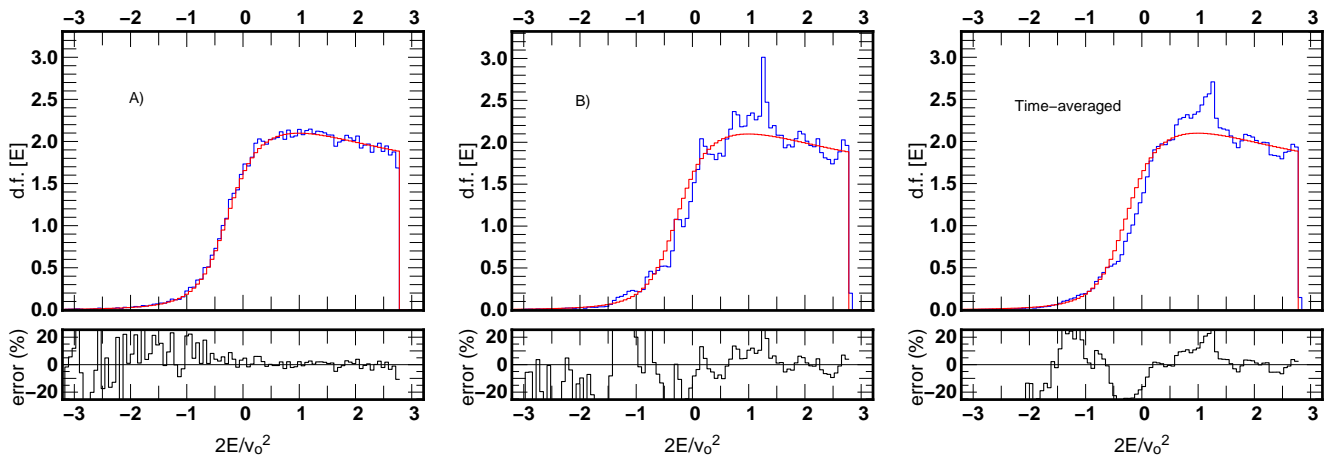

\begin{center}
 \begin{picture}(200,160){
 \put(-190,-120){\epsfig{file=figures/dfE_93k.pdf, width=0.5\textwidth}}
	\put(-20,-120){\epsfig{file=figures/Signal100k_1.pdf, width=0.5\textwidth}}
 \put(150,-120){\epsfig{file=figures/TimeAveraged.pdf, width=0.5\textwidth}}
 }\end{picture}
\end{center}
\caption{Energy distribution for two configurations: A) static black hole at the centre of coordinates; B) black hole on a radial orbit of amplitude $u_o = 0.33$, after 20 time units of evolution. 
In both cases the mass parameter $\tilde{m}_{bh} = 0.3$. 
The energy d.f. averaged over five output times is shown on the right-hand side for comparison. 
The smooth histogram on all three panels was constructed using (\ref{eq:dfelog}). } 
\label{fig:signal} 
\label{fig:timeaveraged} 
\end{figure*}

Figure~\ref{fig:signal} compares the energy distribution function of a set of 93617 orbits when the black hole is set in motion 
with amplitude $u_o = 0.33$, to the initial distribution function, when the black hole sits at the origin. Recall that this latter 
distribution would be time-independent. The orbits were all integrated for $t = 20$ 
unit of times, which corresponds roughly  to 10 revolutions at the edge of the system. The energy axis is once more split 
in 100 bins, and we find once more an rms noise level of $\approx 4\%$. Let us take as 
 one standard deviation a difference  of 5\% with the analytic function (\ref{eq:dfelog}). The signature of black hole motion
 seen on Fig.~\ref{fig:signal}(b) leaves four peaks of more than three standard deviations, and ten with one or more standard deviations, all to the left of $E = -1$.  The most significant peak at $E\simeq 1.31$ has an amplitude of +29\%, 
 or five standard deviations. This corresponds to a circular orbit at radius $u = 1.169$ ($r = u R_c \simeq 0.5355$ units), 
 which is $\sim 2$ times the black hole radius of influence (cf. Fig.~\ref{fig:influenceradius}) and 3.57 times its amplitude of motion, $u_o$. This is clearly an indication of energy exchanges through beat frequencies, i.e., resonances.

 %%tab Table listing the orbital resonances, both Keplerian and Galactic  tab:resonances
\begin{table} 
\caption{Orbital resonances defined in (\ref{eq:commensurate}) for several values of the commensurate ratio $m:n$. 
Results for $m \le n $ (or, ${\cal M} \ge 1$) are labeled Galactic, otherwise they are labeled Keplerian (${\cal M} < 1$). 
\label{tab:resonances}  }
\begin{center} 
\begin{tabular}{cccc}  
Galactic  &  $u$ & $r$ & $2E/v_o^2$ \\ 
$m:n$ & \\
1:1 &  0.748 & &  0.602  \\   
2:3 &  1.292 & &  1.490 \\   
3:5 &  1.479 & &  1.744 \\   
1:2 &   1.849 & & 2.178  \\   
2:5 &   2.390 & &  2.691 \\   
1:3 &   2.918 & &  3.097 \\   
\\
Keplerian  &  $u$ & $r$ & $2E/v_o^2$ \\ 
$m:n $& \\ 
11:10 &  0.663 & & 0.443 \\
6:5    &   0.598 & & 0.317 \\
4:3    &   0.532 & & 0.187 \\
13:9  &   0.489 & & 0.101 \\
3:2    &   0.471 & & 0.064 \\
5:3    &   0.427 & &  -0.031 \\
2:1    &   0.364 & &  -0.171 \\
5:2    &   0.304 & &   -0.320 \\
3:1    &   0.265 & &  -0.433 \\ 
\end{tabular}
\end{center} 
\end{table}

 \subsection{Orbital resonances} 
 To gauge the importance of orbital resonances we identify first the radius and energy of commensurate 
 orbital periods. If we call $\omega^\prime$ the orbital angular frequency of a star on a given orbit of energy $E$, then 
 we need solve the equation 
 
 %%eq Equation for commensurate orbital periods  eq:commensurate
 \begin{equation} 
\nu^{-2}  = \frac{1}{u^2+1} + \frac{\tilde{m}_{bh}}{2u^3}  \equiv \left( \frac{m}{n}\right)^2
 \label{eq:commensurate} 
 \end{equation}
where $\nu = \omega/\omega^\prime$ as in \S1, which requires solution for all  prime  integer ratios $m / n$.  The above equation could be set in terms of ${\cal M}$ defined in (\ref{eq:M}) as done 
 in (\ref{eq:wzero}) in the case of a harmonic potential. Instead we solve for $E,u$ from (\ref{eq:commensurate}), and 
 classify the result as a Keplerian resonance when the corresponding value for ${\cal M } > 2$, which will always be the case when $m > n$, and a Galactic resonance for all other cases $m \le n$. Table~\ref{tab:resonances} lists  the results for 
 a broad range of values of $m:n$. Remarkably, the energy level
  of  {\it Galactic} resonances match almost exactly the
 apparent nodes in the energy distribution function seen on Fig.~\ref{fig:signal}(b). This is in close agreement with our 
 analysis of \S1, when we argued that commensurate values of $\nu$ would give no net work after an integer number of 
 black hole revolutions. This is not the whole picture, however, since the binding energy of individual orbits is constantly changing in time, with as many stars gaining energy as those losing energy. Hence one may think of the energy 
 d.f. of Fig.~\ref{fig:signal}(b) as a standing wave modulated by small sinusoidal modes propagating at a finite
  pattern speed. A clearer picture emerges once we compare Fig.~\ref{fig:signal}(b) to a time-averaged d.f. for the same 
  system. On the right-hand panel of Fig.~\ref{fig:timeaveraged}, we graph the average
   of five energy distributions functions sampled over a time span 
  of $t = 13$ to $17$. This corresponds to $4 / \omega = 4 R_c/v_o \approx 2$ full black hole oscillations. 
  In total 468115 orbits were put to contribution. Comparing this curve to the one displayed on Fig.~\ref{fig:signal}(b), we find fewer  peaks exceeding 
  one standard-deviation.  The smoother appearance of the distribution function  supports 
  the interpretation of sine-like oscillations on Fig.~\ref{fig:signal}(b) as transitory 
  features. Thus we expect phase-mixing to erase such features on a dynamical time-scale. 
   By contrast, the two broad peaks at $2E/v_o^2 \approx 1$ and $-0.5$ remain, their amplitude hardly dented by  the time-averaging. 
  These peaks should, therefore, leave observable features in kinematic- and density maps.

\begin{figure}
\begin{center}
 \begin{picture}(200,250){
\put(-40,-10){\epsfig{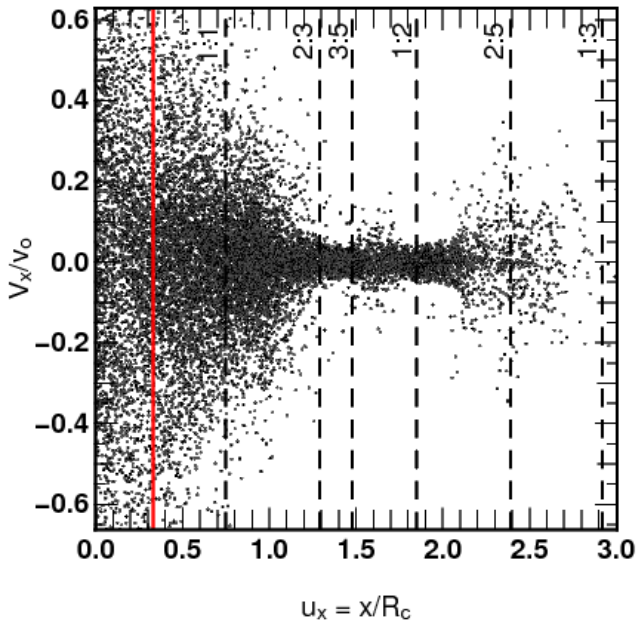}}
 }\end{picture}
\end{center}
\caption{Surface of section in the $y = 0$ plane showing $v_x$ in units of $v_o$ as a function of $u_x = x/R_c$. 
The dashed vertical lines indicate resonances listed in Table~\ref{tab:resonances}.  
 } 
\label{fig:sos100k} 
\end{figure}

 This intuition  is confirmed,  at least partially, 
 by a Poincar\'e section of $(x,v_x)$ in the plane $y=0$. On Fig.~\ref{fig:sos100k} we graph the 
  surface of section of 500 orbits, each evolved for twenty time units. The number of points varies between orbits 
   from $\sim 10 $ and up to $200$, according to the period. We observe large but localised
    scatter in the velocities as a function of $u_x = x/R_c$. The vertical full line at $u_x \simeq 0.3$ 
    indicates the amplitude of motion of the 
    black hole, while the dash lines are the location of resonances listed in Table~\ref{tab:resonances}. The scatter decreases   rapidly as we move to large radii. Specifically there is a sharp drop as we reach beyond the 1:1 resonance, and thereafter
    significant scatter is centered around $u_x \approx 1.6$ and 2.5. These values of $u_x$ 
    correspond to energies of $2 E/v_0^2 \approx 1.9 $ and 2.8, respectively, matching the features seen on the energy 
    distribution function (cf. Fig.~\ref{fig:timeaveraged}). The broad peaks seen on graphs of the energy 
    d.f. falls inside the 1:1 resonance and are lost in the scatter on Fig.~\ref{fig:sos100k}.

 %%fig   fig:sos100k

 \subsection{Surface density, velocity maps}
Maps of the surface density and velocity field are of interest. The configurations are isotropic 
 initially when the black hole starts off at the centre of coordinates. At later stages 
 neither the density nor the velocity fields  respect this initial property. We toyed with 
 the idea of plotting both surface density $\Sigma$ and velocity field in cylindrical 
 coordinates centered either on the system's centre of mass, or the black hole. This turned
 out to be useful only when both coincides, which will only occur when the black hole is 
 on a radial orbit. Instead, we opted to map out both quantities on  a uniform Cartesian grid. 
 This has the advantage of an unbiased linear resolution over all space and is identically suited 
 to any type of black hole orbit (radial, circular or other). The surface density is obtained 
 at any time by a simple count-in-cell (CIC)  technique. No smoothing or averaging of neighbouring 
 cell has been performed. Density profiles and their significance  should be interpreted with due consideration to 
 root-n noise: a typical grid would have $30\times 30$ mesh points, and hence a mean count per 
 cell of at least $\gtabout 4\,10^4/900 \simeq 44$, which translates to relative fluctuations of 15\%. In practice 
 we have used on order $10^5$ orbits, and so the noise level always falls in the range 10\% - 15\%. As we will 
 see, the density fluctuations that we measured at times exceeded 60\% of the reference initial profile, giving  a 
 signal-to-noise ratio of at least 4 and perhaps as high as 7. 
 
 \subsubsection{Flat-fielding the velocity map}
 The initial velocity (\ref{eq:logvc}) is known at any point in space, however it is set in the centre of mass reference 
 frame. Since we wish to map out velocities on a Cartesian grid, we must be cautious to compute the expected mean 
 velocity in any cell for comparison. This poses a problem around the origin of coordinates, when the radius $\sim$ a 
 few grid cells only. Calling $\delta u$ the grid size, we find from (\ref{eq:logvc}) an absolute error on $v_c$ of 
 
 %%eq Relative error on the circular velocity due to grid resolution issues  eq:dv 
 \begin{equation}
 \frac{|\delta v_c|}{v_c} \le \left( \frac{|\delta u|}{2u}\right) \left| \frac{2}{1+u^2} + \frac{\tilde{m}_{bh}}{2\,\max\{u,\varepsilon\}}\frac{1+u^2}{u^2} \right| 
 \label{eq:dv}
 \end{equation}
 which becomes large when $u \ltabout \delta u$. We then compute (\ref{eq:dv}) for each orbit falling inside a given 
  mesh and take the average square difference with the local circular velocity: 
  
%%eq Definition of the relative dispersion sigma at time t on a grid   eq:sigma 
\begin{equation}
\frac{1}{n}  \sum_{i=1}^n \left( \frac{ ||v||- v_c }{1 + |\delta v_c|/v_c} \right)^2 \equiv \sigma^2 
  \label{eq:sigma}
\end{equation}
where the sum is over all $n$ orbits  inside the mesh at time $t$ obtained by CIC. This gives a direct measure of 
the local dispersion as a result of black hole motion, and a reference map to eliminate noise when the black hole is 
fixed. For that case, we find using (\ref{eq:sigma}) residual errors not larger than $1:10^4$ or 0.01 \%. 

\subsubsection{Time sequence} 
We graph on Fig.~\ref{fig:density} the time-sequence of the surface density and dispersion $\sigma$ for 93607 orbits 
integrated over 20 time units. The figure shows two sets of two rows,  regrouped to help match features seen in the 
density, to those seen in de velocity field. Initially the orbits are isotropic and circular, which explains the two feature-less
frames at the top-left corner of Fig.~\ref{fig:density}. At subsequent times, the plots show very pronounced and 
fast-evolving features, both in maps of the density and velocity. The scale of surface density was chosen to identify 
peak density enhancements of 60\% when compared to the initial profile. The core radius $R_c \approx 0.45$, nearly twice  
the black hole influence radius of $\simeq 0.56 R_c$, is displayed as the dashed circle. Inside and up to that radius, the 
 mass profile shows arcs, bubbles and other transient features, all suggestive of unsteady, perhaps chaotic, orbital 
 motion. Outside that radius, we find more steady, ring-shaped features which match the position of resonances (Table~\ref{tab:resonances}). 
 
By comparison, the relative velocity dispersion   peaks at $\approx 25\% $ 
 of the local circular velocity $v_c$. Not surprisingly, the largest deviations in velocity nearly always coincide with the 
 position of the black hole (red dots on Fig.~\ref{fig:density}). The most remarkable features on these frames are the
 large dispersions measured well outside the core radius. A particularly striking sequence runs from $t = 15$ to 17, 
 when a large arc seems to close up on itself in the region $x \approx -0.75, y \simeq 0$. This large dispersion is found at 
 a distance some 5 times larger than the amplitude of the black hole's orbit. If we scaled these features to the MW, then 
 an anomalous local velocity dispersion would be expected up to from 3 to 4 pc away from Sgr A$\star$. 
In conclusion, the black hole has a very strong impact on circular orbits, both in terms of spatial features and kinematics,
 up to $\approx 2$ to 3 times its  radius of influence (see also Fig.\ref{fig:sos100k}). 
  
%%fig Maps of the surface density  fig:density
\begin{figure*}
\begin{center}
 \begin{picture}(200,570){
%% Density : 
%    \put(-50,20){\epsfig{file=figures/astroph_9.png, width=0.7\textwidth}}
%    \put(-150,420){\epsfig{file=figures/Density100k_0.png, width=0.3\textwidth}}
%    \put(25,420){\epsfig{file=figures/Density100k_13.png, width=0.3\textwidth}}
%    \put(200,420){\epsfig{file=figures/Density100k_14.png, width=0.3\textwidth}}
% \put(-150,295){\epsfig{file=figures/Vfield100k_0.png, width=0.3\textwidth}}
% \put(25,295){\epsfig{file=figures/Vfield100k_13.png, width=0.3\textwidth}}
% \put(200,295){\epsfig{file=figures/Vfield100k_14.png, width=0.3\textwidth}}
%%
%	\put(-150,130){\epsfig{file=figures/Density100k_15.png, width=0.3\textwidth}}
% 	\put(25,130){\epsfig{file=figures/Density100k_16.png, width=0.3\textwidth}}
% 	\put(200,130){\epsfig{file=figures/Density100k_17.png, width=0.3\textwidth}}
%\put(-150,5){\epsfig{file=figures/Vfield100k_15.png, width=0.3\textwidth}}
% \put(25,5){\epsfig{file=figures/Vfield100k_16.png, width=0.3\textwidth}}
% \put(200,5){\epsfig{file=figures/Vfield100k_17.png, width=0.3\textwidth}}     
    
 }\end{picture}
\end{center}
\caption{Time-sequence showing the surface density $\Sigma$ and flat-fielded velocity dispersion $\sigma$ for a system with an $\tilde{m}_{bh} = 0.3 $ mass black hole on a radial orbit of amplitude $u_o = 0.33\, (R_o \approx 0.15)$. The frames show maps of $\Sigma$  (first and third rows) and 
dispersion $\sigma$ (second and fourth rows) at times $t = 0, 13, 14, 15, 16 $ and $17$ units. The 
black hole is shown as a red dot, while the cross is the origin of coordinates. The black arrows are the local net  angular momentum, which 
remains at the root-n level everywhere. 
}
 \label{fig:density} 
\end{figure*}
 
 %%fig Profile of the mean square vel dispersion  (1d)  fig:sigma1d
\begin{figure*}
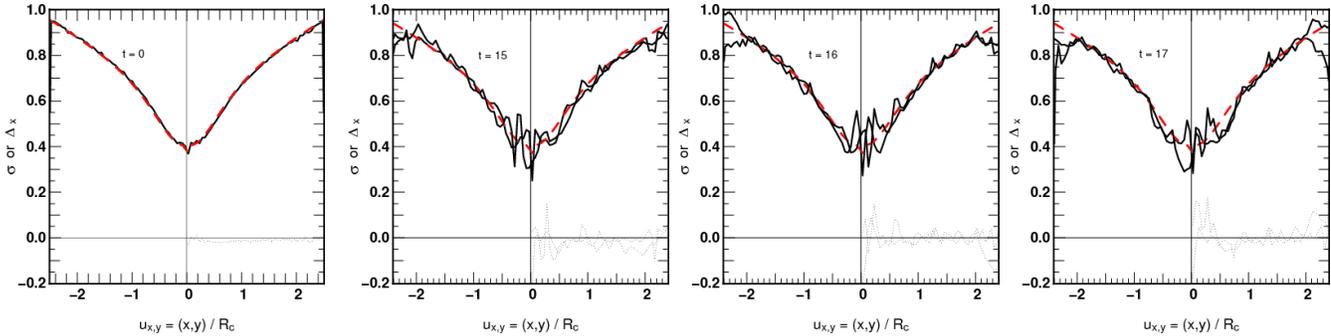

\begin{center}
 \begin{picture}(200,150){
% Density : 
\put(-155,0){\epsfig{file=figures/sigma1d_00.png, width=0.24\textwidth}}
    \put(-25,0){\epsfig{file=figures/sigma1d_15.png, width=0.24\textwidth}}
    \put(100,0){\epsfig{file=figures/sigma1d_16.png, width=0.24\textwidth}}
    \put(225,0){\epsfig{file=figures/sigma1d_17.png, width=0.24\textwidth}}
  }
  \end{picture}
\end{center}
\caption{Time-sequence showing the line-of-sight root-mean-square velocity dispersion, $\sigma$, for the reference case of $\tilde{m}_{bh} = 0.3 \simeq u_o$.  
A total of 100 bins were used for two  viewing angles, down the x- and y-axis, in each case. 
At $t = 0 $ (panel to the extreme left) the analytic expectation with a black hole at the centre of coordinates is well recovered from 69773 circular orbits. The dashed line shows the theoretical 
expectations which drops to zero at the origin (all motion is orthogonal to the line-of-sight). The 
solid line gives the dispersion recovered from summing over all orbits. 
When $t > 0$, $\sigma$ differs from the expected value up to $u_{x,y} \simeq u_o$, the black hole 
amplitude of motion. At large distances, the broken curve suggests Poisson noise from small-N statistics. The thin dotted lines are the differences about the coordinate centre $\sigma(u) - \sigma(-u)$ for each line-of-sight. 
}
 \label{fig:sigma1d} 
\end{figure*}

 %%fig Profile of the mean velocity (1d)  fig:meanv1d
\begin{figure*}
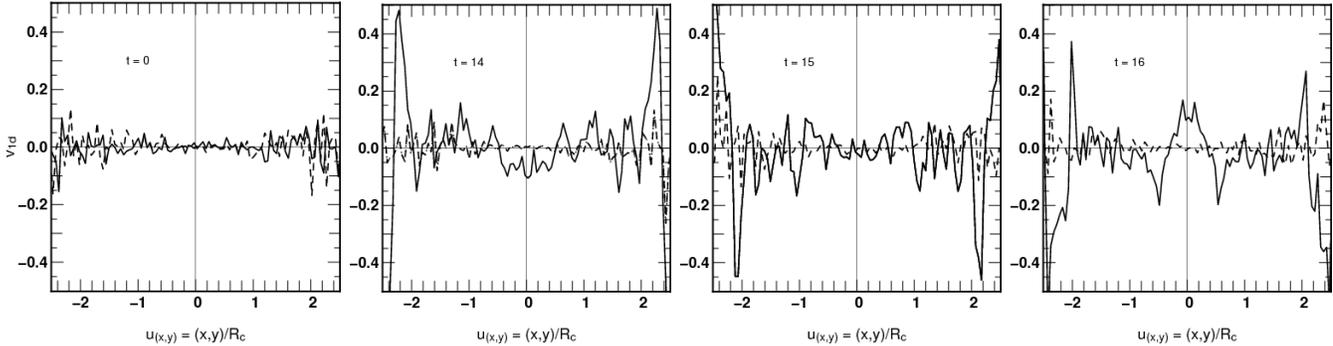

\begin{center}
 \begin{picture}(200,150){
% Density : 
\put(-155,0){\epsfig{file=figures/meanv1_t0.png, width=0.25\textwidth}}
    \put(-25,0){\epsfig{file=figures/meanv1_t14.png, width=0.24\textwidth}}
    \put(100,0){\epsfig{file=figures/meanv1_t15.png, width=0.24\textwidth}}
    \put(225,0){\epsfig{file=figures/meanv1_t16.png, width=0.24\textwidth}}
  }
  \end{picture}
\end{center}
\caption{Time-sequence showing the line-of-sight mean velocity for the reference calculation with black hole mass 
$\tilde{m}_{bh} = 0.3$ and amplitude $u_o = 0.33$. The velocity was measured when looking down the $x$-axis 
(solid line) and the $y$-axis (dashed line). The $x$-axis data show cyclical variations with a peak signal-to-noise ration 
inside $u = 1$ (or, $r = R_c$).  
}
 \label{fig:meanv1d} 
\end{figure*}

 \subsection{Projected velocity field}
 The line-of-sight velocity offers a direct way to measure the effect of black hole motion. 
  Our strategy, then, is to apply a slit of some $100$ mesh running across the 
 $x$- and $y-$axis (viewing angles of 0$^o$ and 90$^o$ to the black hole orbit, respectively), 
 and compute both the integrated one-dimensional velocity and the square velocity dispersion at each 
 mesh point. The expected velocity in projection   $v_{1d}(x\ {\rm or}\ y) = 0$ for isotropic initial conditions. 
 The analytic calculation of the dispersion $\sigma_{1d}$ 
 proceeds straightforwardly from (\ref{eq:logrho}) and (\ref{eq:logvc}). Inspection 
 of several time frames, using either $x$ or $y$ as the line-of-sight, revealed no significant deviation from the 
 profile derived for circular motion, at any point of evolution. Specifically, we looked for asymmetries in the dispersion when comparing the two 
 viewing angles, but found none (to within $\approx 2 \times$ the noise level). Consequently we discarded runs of $\sigma_{1d}$  from analysis. (In \S5 we comment on the dispersion in the context of  fragmentation modes.)
  
 The situation with  $v_{1d} $ is more profitable. Fig.~\ref{fig:meanv1d} graphs the line-of-sight velocity 
 for three time frames of the reference C3 calculation with 
 $\tilde{m}_{bh} = 0.3$  and amplitude $u_o = 0.33$. The results are shown 
 for two viewing angles in each case at times $t = 0, 14, 15 $ and $16$. A total of 83697 orbits were binned
  on a mesh of 100 points. The Poisson fluctuations $\sim 3\%$ on average but are less than $1\%$ inside $|u| < 1$, and 
  $\approx 10\%$ near the end-points due to sampling effects. 
 Looking down the $y$-axis orthogonally to the black hole orbit, we find $v_{1d} = 0$ everywhere 
 to root-n noise, a result which confirms the intuition that the $y$-component of the velocity field preserves the 
 initial symmetry through the $z-x$ plane (Fig.~\ref{fig:meanv1d}, dashed lines). A sample  of these three and two more frames
 gave very similar results for the root mean-square scatter $\sqrt{<\!\delta v_{1d}^2\!>}  \simeq 0.002$ when averaged over the mesh. 
 
 When we switch lines-of-sight, we find large time-dependent oscillations of  $v_{1d}$ of a half-period $\simeq 2$, close to half the 
 orbital period $= 2\pi R_c/v_0 = 2.9$ of the black hole (Fig.~\ref{fig:meanv1d}, solid line). 
 This highlights a close relation between the  stars' 
 angular momentum $L_z = xv_y - yv_x$ and the phase of the black hole orbit: the torque $\Gamma_z \approx  
 GM_{bh}/r^2\, (y_\star/r) (x_\star - x_{bh}) \hat{z}$ (where $r = ||\bmath{r}_\star - \bmath{r}_{bh}||$ is
 the distance between the star and the black hole) will be positive for half the stars inside a given mesh centered on $y_\star$. This torque, then, will give a boost to the momentum of stars when  $\Gamma_z \cdot L_z > 0$  (otherwise the torque will oppose
 the star's momentum,  and decrease its magnitude). As a result  the line-of-sight
  velocity does not integrate to zero and 
 shows cyclic variations with $u_y$. Inspections of the same five frames as before 
 using the $x$-axis as line-of-sight showed that 
  $|v_{1d}| $ varied from $0.008$ to $0.034$ with a root mean square scatter $\sqrt{<\!\delta v_{1d}^2\!>} \approx 0.03$. The 
  run from $t = 14$ to $16$ is an example of a sequence during which $|v_{1d}|$,  averaged over all bins, goes from 0.032, to a minimum 0.004, and then back to 0.034. We stress that the trend with 
  time  is averaged over all bins; it is therefore of a much larger amplitude
   than the scatter seen inside $|u_y| = 1$. Alternatively, we may compute the ratio of
    $|v_{1d}|  $ measured down one axis, to the values obtained for the initial configuration at $t=0$. We find a scatter $\sqrt{<\!\delta v_{1d}^2\!>} \approx 0.015$ averaged over all bins, but only $\approx 0.005$ inside $|u| = 1$ owing to better statistics. 
    If we refer to these 
    data as noise, then the signal-to-noise ratio  when $|v_{1d}|  $ goes through
    a maximum reaches 10 near $u \approx 0$ at these times. 
    The rms scatter of $x$-axis data is systematically $\approx 3\times$ higher that of 
    $y$-axis data, at all times (Table~\ref{tab:results}). 
    
    The impact of black hole motion on the distribution of angular momentum $L$ appears clearly on a graph of the momenta distribution function. The panel to the top left on Fig.~\ref{fig:L} (labeled `cold') shows the d.f. at time $t = 17$ for the reference calculation, compared to the initial profiling (thin dashed line). 
    The angular momenta are distributed in a highly symmetrical fashion about the $L = 0$ axis. Each 
    feature marking a departure from 
    the initial distribution is matched pair-wise for the same value of $|L|$. Thus the sum of all stellar momenta remains 
    constant. Fig.~\ref{fig:L} (top, left-hand panel) illustrates the r\^ole of the black hole acting as a catalyst to transfer angular momentum, as well  as energy, to the stars. 

%%tab List results conditions tab:results
\begin{table} 
\caption{Results for the models listed in Table~1. \label{tab:results}  }
\begin{center} 
\begin{tabular}{lcccc}  
\multicolumn{5}{c}{Circular stellar orbits (`cold')} \\ 
Name  & rms $(2\delta E/v_o)^2$ &\multicolumn{2}{c}{rms $(\delta v_{1d}^2)/v_o$} & Comment \\ 
            &              $\pm 4 [\%]$  &  x-axis       & y-axis    &        \\
C1        &               $0 $     &    0.015              &  0.015             &     $t = 0$ data \\
C2        &           $12 \pm 2$   &    0.022              &  0.020     & \\
C3        &         $29.2 $   &    0.071 $\pm 0.01$   &  0.042 &  Ref. case\\
C4        &         $36.9 $   &    0.044              &  0.026 &    \\
C5        &         $51 $     &    0.038              &  0.024 &     \\
\\
C2s     &           $24 $     &    0.025              &  0.021       &  Shadows C2     \\
C3s     &           $50 $     &    0.041              &  0.022       &  Shadows C3    \\
\\
\multicolumn{5}{c}{`Warm'  or `Hot' runs } \\ 
Name  & rms $(2\delta E/v_o)^2$ &\multicolumn{2}{c}{rms $(\delta v_{1d}^2)/v_o$} & Comment \\ 
            &              $\pm 4 [\%]$  &  x-axis       & y-axis    &        \\
W1       &           $18.8 \pm 1.6$       &  0.050  & 0.015 &   \\
W1c      &           $22.0 $              & 0.041   & 0.031  & circ. BH orbit \\
H1        &          $28.5 \pm 3.5$       & 0.034   & 0.014 &  \\
\\
W2       &            $32 \pm 1 $         & 0.033   &  0.016 & \\ 
\end{tabular}
\end{center} 
\end{table}

\subsection{Exploring different configurations}
 \subsubsection{Changing the velocity field}
Our choice of circular orbits has some bearing on the outcome of the calculations. In this section we  
assess to what extent  features seen  on  Fig.~\ref{fig:density} are specific to our 
choice of initial conditions. 

To that end, we first took the same set of orbits but modified initial velocities by $\pm 10\%$ in 
magnitude,  so that the orbits were no 
longer circular initially. This exercise produced similar features in the maps of density and velocity as for the reference setup, 
a hint that the initial response of the stars, and the general features, are not sensitive to imposing strict circular motion to 
the initial conditions. We then toyed with the idea of computing a velocity field consistent with the logarithmic potential 
based on moments of the Boltzmann equations (e.g., Binney \& Tremaine 1987). Doing so however would have 
meant introducing a new energy d.f. and made comparisons with the case of circular orbits more difficult. Instead, we opted to 
keep the same energy d.f. and to change the velocity field by reorienting the velocity vectors randomly inside some angle chosen in the interval 
$\pm \theta$; here $\theta = 0$ gives the original d.f. with all orbits circular. 
Two new configurations were setup, one with $\theta = \pi/4 $ (giving a cone of opening angle 45$^o$), which 
we label {\it warm}, and a second with $\theta = \pi $ (fully random in azimuth), which we label {\it hot}. The distributions were
otherwise unchanged from the reference model C3. The full list of model parameters is given in Table~\ref{tab:initialconditions}. 

%%fig Figure showing sequence of Cold, Warm Hot runs fig:sequencecwh
%
\begin{figure*}
\begin{center}
 \begin{picture}(200,600){
%  \put(-50,43){\epsfig{file=figures/astroph_12.png, width=0.8\textwidth}}
% \put(-150,443){\epsfig{file=figures/L-cold.png, width=0.3\textwidth}}
% \put(15,440){\epsfig{file=figures/Rho_Cold.png, width=0.35\textwidth}}
% \put(200,440){\epsfig{file=figures/v1d_Cold.png, width=0.3\textwidth}}
%	\put(-150,293){\epsfig{file=figures/L-warm_1.png, width=0.3\textwidth}}
% 	\put(15,290){\epsfig{file=figures/Rho_Warm.png, width=0.35\textwidth}}
% 	\put(200,290){\epsfig{file=figures/v1d_Warm.png, width=0.3\textwidth}}
% \put(-150,148){\epsfig{file=figures/L-hot.png, width=0.3\textwidth}}
% \put(15,145){\epsfig{file=figures/Rho_Hot.png, width=0.35\textwidth}}
% \put(200,145){\epsfig{file=figures/v1d_Hot.png, width=0.3\textwidth}}
%	 \put(-150,0){\epsfig{file=figures/L-warmcirc.png, width=0.3\textwidth}}
%	 \put(15,-3){\epsfig{file=figures/Rho_CircBH.png, width=0.35\textwidth}}
%	 \put(200,-3){\epsfig{file=figures/v1d_CircBH.png, width=0.3\textwidth}} 
 }\end{picture}
\end{center}
\caption{This graphs from left to right: the angular momentum distribution, the surface density, and the line-of-sight velocity 
for four models. The top row gives the results for the reference C3 calculation; the second row is the `warm' W1 calculation; 
and the third row is the case H1 of a `hot' distribution (see Table~\ref{tab:initialconditions} and \S4.5 for details). 
All quantities were analysed after $t = 17$ time units of integration; a minimum of 93 000 orbits were used to sample each of
 the parameters. The last row 
at the bottom gives the solution for the case W1c of a black hole set on a circular orbit. Note the strong $m=0$ mode on the
 map of the surface density. 
 } \label{fig:sequencecwh} 
%%fig Histograms of L  fig:L
 \label{fig:L} 
 %%fig Histograms of L  fig:L
 \label{fig:densitycwh} 
\end{figure*}

The results are summed up graphically on Fig.~\ref{fig:densitycwh}. The top three rows show the angular momentum d.f. (left-hand panels), the surface density (middle) and $v_{1d}$ for two projection axis (right-hand panels) in turn for the `cold' configuration of circular orbits, the `warm' and `hot' initial conditions. All data shown on the figure were taken at time $t = 17$ so the black hole assumes the same position and velocity in each case. The rings seen in the surface density map of the cold run outside 
$r = R_c \approx 0.5$ (shown as a dashed circle) 
have disappeared, and only the strongest feature inside $R_c$ remains visible for warm and hot 
initial conditions. Thus the filamentary structures seen on Fig.~\ref{fig:density} are attributable to the strong response from 
circular and near-circular orbits. These would make up a small fraction of warm and hot distributions. In these two cases, the 
number of low-$L$ orbits is higher and more stars visit the central region on eccentric orbits. These scatter 
off the black hole effectively and acquire large angular momenta, a feature which can be measured up from the 
swelling of the momentum d.f. 
for  large $|L|$ and the depletion of the d.f. around $L = 0$ (Fig.~\ref{fig:sequencecwh}, left-hand panels, 2nd and 3rd row). 
The black hole motion has $u_o = 0.15$ translating to $r =  0.069$ and a radius of influence $\simeq 0.23$. Therefore all stars 
with an initial angular momentum lower than $r\times v \approx 0.34$ would come within a distance $\ltabout r_{bh}$ 
of the black hole 
in one revolution\footnote{We have set $r = 0.23 + 0.069 $ or $u = 0.65$ and $v(u)$ from (\ref{eq:logvc}).}.
Fig.~\ref{fig:sequencecwh} shows this estimate in good agreement with the numerical computations. A fraction of 
stars concerned is 
 about 12\% in the `warm' calculation, but close to 25\% in the `hot' run, when circular orbits have all but been wiped out. 

Although details of the density and angular momentum profiles are much affected in the new configurations, 
compared to the reference cold one, the same does not 
apply to the line-of-sight velocity. The response of stars for both warm and hot computations show 
a distinct signature of black hole motion in the sense that once again $v_{1d}$ fluctuates significantly more when 
measured down the  axis parallel to the 
black hole's orbit. The averaged scatter in both the cases compares well with the data for the 
cold configuration of circular motion, at least in the interval  from $u = 0$ to $|u| = 1$, or roughly two times the black hole radius of influence 
 (Fig.~\ref{fig:sequencecwh} and Table~\ref{tab:results}). In conclusion, the black hole motion still 
 imprints the kinematics of both
 warm and hot configurations, albeit to a lesser degree than for the case when stars are on circular orbits. 

 %%fig Profile of the energy d.f. for four configurations  fig:compareE
\begin{figure*}
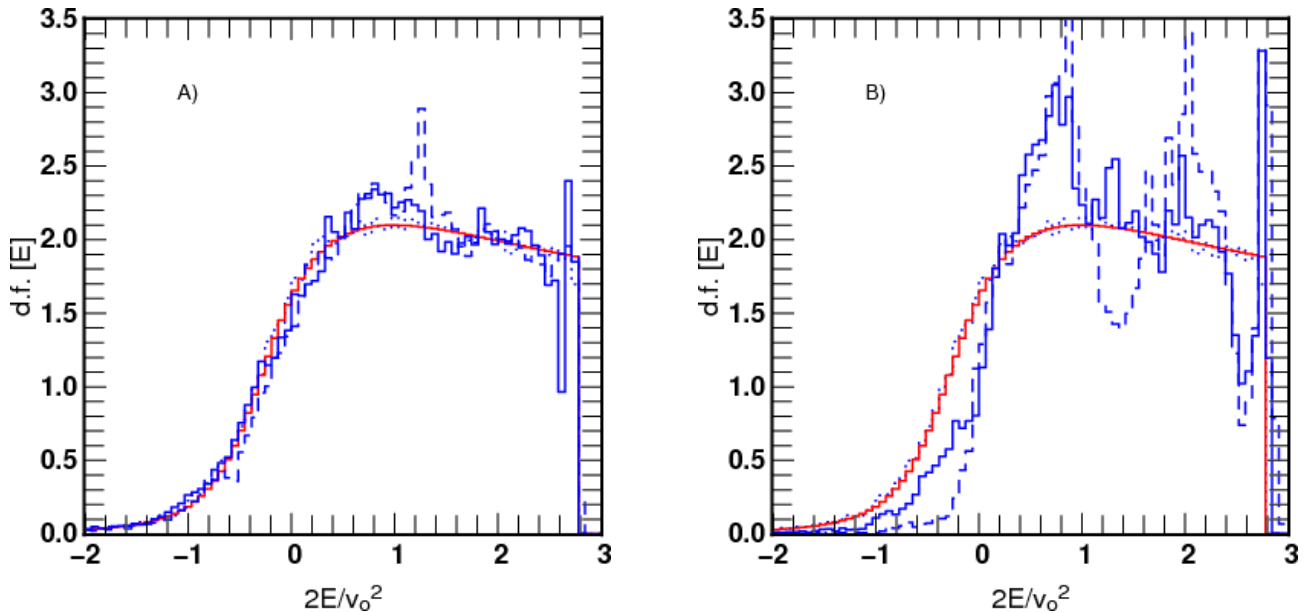

\begin{center}
 \begin{picture}(200,250){
% Density : 
\put(-135,0){\epsfig{file=figures/compareE_a.png, width=0.45\textwidth}}
    \put(125,0){\epsfig{file=figures/compareE_b.png, width=0.45\textwidth}}
  }
  \end{picture}
\end{center}
\caption{
This graphs the energy distribution for four values of the  black hole's amplitude of motion $u_o = R_o /R_c$. The dimensionless mass parameter $\tilde{m}_{bh} = 0.3$ in all the cases. 
On panel (a)  we plot two solutions with $u_o = 0.07 $ (solid line) and $0.15$ (dashed line); on (b) we set $u_o = 0.21 $ (solid) and $0.30$ (dashed). The smooth curve is the analytic solution (28). The dots shows a discrete realisation with 69977 orbits and $u_o = 0$ (axially symmetric potential). 
}
 \label{fig:compareE} 
\end{figure*}

\subsubsection{Changing the black hole parameters: scaling}
Equations (\ref{eq:network2}), (\ref{eq:logmass})  and (\ref{eq:mbhtilde}) may be combined to give a proportionality 
relation between the work $W$ and black hole amplitude of motion and mass function ${\cal M}$. Keeping only the 
radius-dependent terms we find 

%%eq Rough scaling relation between work and black hole orbital parameters. 
%
\begin{equation}
W \propto \frac{u_o}{u^2} \left[ \tilde{m}_{bh} + \frac{u^3}{u^2+1} \right]  
\label{eq:scaling}
\end{equation}
so that the ratio $W/(u_o \tilde{m}_{bh})$ is roughly homogeneous in $1/u^2$ whenever $u^3 \ll 1$. This limit would allow 
to retrieve a scaled version of any calculation following a redefinition of the black hole mass and/or amplitude of motion 
by a suitable rescaling of the lengths. However the limit $u \ll 1$ implies that the orbit is well inside the black hole radius 
of influence. Only a very small fraction of orbits will be found there. Nevertheless, (\ref{eq:scaling}) suggests that two 
configurations with $u_o \tilde{m}_{bh} $ kept constant would yield a similar net work $W$ on some orbits and so 
possibly the same or comparable imprint on the stellar energy d.f.

We investigated this with two configurations, C2s and C3s, tailored to shadow runs C2 and C3 (cf. Table~\ref{tab:initialconditions}). The two `shadow' runs both had a black hole mass equal to half that of C2 and C3, however 
twice the amplitude of motion $u_o$. 
 Table~\ref{tab:results} lists the root mean square deviations of the energy d.f. compared to the analytic curve (\ref{eq:dfelog}).
The impact on the energy distribution function is clearly stronger for the large-amplitude 
black hole runs. The energy deviations of $24\%$ 
are nearly as large for an $\tilde{m}_{bh} = u_o = 0.15$ configuration as 
those obtained with the same configuration  but with $\tilde{m}_{bh} = 0.30$, for which we found $29\%$ deviations on the mean.  
In practice, we find that the energy fluctuations scale almost linearly with the black hole's orbital radius $u_o$, a conclusion 
which we  reached by comparing the results for cases C1 to C5. What is more, the amplitude of the response of the stars appears 
 robust to details of the velocity field, as seen when comparing 
configurations with warm or hot stellar velocity fields (W1, H1 and W2). A configuration with a black hole set on a circular orbit 
did not produce significantly different results compared to similar configurations with the same mass and amplitude $u_o$ 
(case W1c, Table~\ref{tab:results}). Taken together, these results highlight the importance of the 
black hole effective orbital cross section $\propto u_o^2$ in transferring binding energy to the stars. 

%%fig Maps of the Toomre number @ three different times for an circular motion setup fig:jeans
%
\begin{figure*}
\begin{center}
 \begin{picture}(200,160){
 \put(-100,0){\epsfig{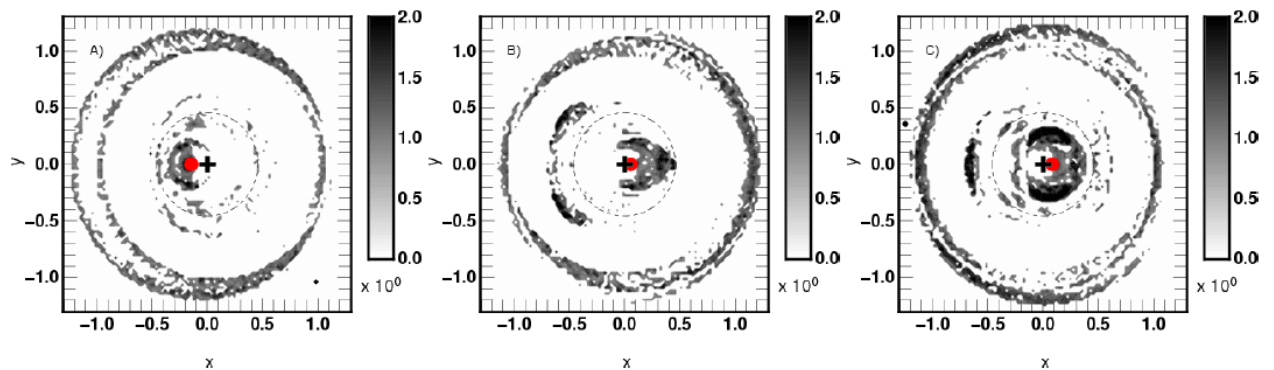}}
% \put(-160,0){\epsfig{file=figures/Jeans_t15.png, width=0.35\textwidth}}
%	\put(20,0){\epsfig{file=figures/Jeans_t16.png, width=0.35\textwidth}}
% \put(200,0){\epsfig{file=figures/Jeans_t17.png, width=0.35\textwidth}}
 }\end{picture}
\end{center}
\caption{Maps of the Toomre number $Q_J$ for the reference calculation C3 at A) t = 15; B) t = 16 and C) t = 17 time units of 
evolution. The shade indicates stability to fragmentation modes. The white regions have $Q_J < 1$ and would be 
susceptible to fragment.} 
\label{fig:jeans} 
\end{figure*}

\section{Discussion} 
Oscillations of a massive black hole about the centre of a host galaxy leave a signature on 
the kinematics of surrounding stars. We showed from an analytic harmonic 
potential model that stars on circular orbits lose or gain energy according to the relative phase 
between their and the black hole's orbit: for an evenly sampled distribution function, half of the 
stars gain energy, the other half loses energy. The black hole plays the role of a catalyst 
by allowing energy exchanges between the stars. 

We explored a range of orbits with black hole and stars in a logarithmic potential with 
an Bulirsch-Stoer  numerical integration scheme (Press et al. 1992) using on the order of $10^5$ orbits.  
The feedback of stars on the black hole orbit was neglected, an approach motivated by 
the large black hole to stellar mass ratio. 
With stars set on circular orbits, we found a strong response to black hole motion, 
at distances ranging up to 3 or 4 times the black hole's radius  of influence (see Fig.~\ref{fig:poincare} and \ref{fig:density}). Such a strong response was also seen in perturbed circular orbits, with velocity perturbations in the range $\pm 10\%$. We quantified the impact of 
black hole motion on the stars' energy distribution function. We measured root-mean square 
deviations growing linearly with the amplitude of motion, $u_o = R_o/R_c$, where $R_c$ is 
the core-radius of the logarithmic potentials. We obtained a significant response (signal-to-noise ratio $> 4$) even for modest amplitudes of $\sim r_{bh}/3 $, where $r_{bh}$ is the black hole's radius of influence
(cf. Eq.~\ref{eq:rbh} and Table~\ref{tab:results}). Worried that these results concerned circular or near-circular orbits
only, we reset the velocity field (`warm' and `hot' configurations, see  \S4) but kept the 
original energy distribution function unchanged. This resulted in washed out features in 
maps of the surface density and velocity dispersion (see Fig.~\ref{fig:sequencecwh}) but 
left a signature on the energy d.f. of the same magnitude, with once again the black hole 
amplitude of motion the main agent responsible for re-shaping the energy d.f. (Table~\ref{tab:results}). 
An analysis of the angular momentum d.f. showed that non-circular orbits visit the centre 
more frequently and couple more strongly with the black hole at some point on their orbit. 
Thus, although the energy exchange mechanism identified for circular orbits plays a secondary 
role in systems with warm and hot velocity fields, the black hole motion still induced strong 
anisotropy. This was most clearly seen when profiling the net line-of-sight velocity $v_{1d}$
 resulting from 
two different projection angles (see Fig.~\ref{fig:sequencecwh}, right-hand panels). 
 Any level of anisotropy is attributable to the motion of the black hole, since a static hole 
 would have left the initial isotropic d.f. unchanged. 

\subsection{Comparison with MW data}
The largest values of $v_{1d}$ were obtained from a viewing angle parallel to the motion of 
the black hole. 
Contrasting these values to the root mean square velocity dispersion, we find a ratio of 
$<\!|v_{1d}|\!>/\sqrt{<\!v_{1d}^2\!>}  \approx 25\% $ at maximum value, inside the hole's radius 
of influence (see Fig.~\ref{fig:meanv1d} and \ref{fig:sequencecwh}). 
Applying  this to Milky Way data, where the mean velocity dispersion rises to $\sim 180$ km/s inside 1 pc  of Sgr A$\star$ (Genzel et al.1996)
 we obtain streaming velocities in the range $\sim 40 $ km/s, 
a rough match to the values reported recently by Reid et al. (2006). The surface density profile 
  shows a break at radius $r_{br} \sim 0.2 $ pc (Sch\"odel et al. 2007). Inside $r_{br}$, the 
  volume density is fitted with a power-law index $\gamma \simeq 1.2 $ which falls outside the 
  range 3/2  to 7/4 of the Bahcall-Wolf solution. Black hole motion of an amplitude 
  $R_o \sim r_{br} $ might cause such a break. The ratio $r_{br}/r_{bh} \sim 0.2 $ compares well with the value $\approx 0.3$ adopted for our reference calculation (Table~\ref{tab:initialconditions}). 
  % These values should be interpreted with caution since the calculations 
  % presented here did not take into account all the complexity of the MW centre dynamics. 
  
  \subsection{Circular black hole orbit}
  We worried that a black hole set on a 
  radial orbit might trigger only a 
  subset of resonant modes from the stars, in contrast to the more probable 
  situation where  the  hole's orbit has a finite angular momentum. 
  Recall the analysis of \S2.2, where the response was 
  stronger for aligned orbital angular momenta. 
  To test this idea, we re-ran the `warm' W1 calculation with the black hole now set on 
  a circular path at a radius $u_o = R_o/R_c = 0.097$, or $\simeq 50\% $ 
  its radius of influence (cf. W1c, Table~\ref{tab:initialconditions} and \ref{tab:results}). 
    As the black hole orbits the centre, an $m = 0$ density mode  develops which 
    shows up as a trailing arm on Fig.~\ref{fig:sequencecwh}, bottom row, middle panel. 
    The 
    black hole orbit is anti-clockwise. The arm stretches radially 
    from 1 to $\approx 2$ times $r_{bh}$. Its integrated mass $ \simeq 40\%$  
    the black hole's mass, and so if the gravity of the arm were taken into account, 
    the torque that this would produce would modify the black hole's 
    orbit significantly, an effect which was neglected here. 
    The averaged line-of-sight velocity of the stars, on the other hand, showed  
    spatial variations of the same amplitude as in the other cases with a strictly radial 
    black hole orbit of a similar amplitude. This  
    result comforts the thought that black hole motion may yet give rise to an observable 
    kinematic signature (especially in the profile of $v_{1d}$), 
    regardless of the precise parameters of its orbit.
  
 \subsection{Jeans instability} 
Our approach suffers from a severe limitation, in that it does not integrate the full response 
of the stars to their own density enhancements. These could become bound structures
which would alter the dynamics globally. To inspect whether this could have an 
influence over the evolution of the velocity field, we computed the Toomre number 

\[ 
Q_J \equiv \frac{\sigma\Omega}{G\Sigma} = \frac{\sigma^2}{G\Sigma {\rm d}l} 
\]
at each mesh points of the simulations space.  Here the mesh size ${\rm d}l \simeq 0.02$,   
and the velocity dispersion $\sigma$ given by (\ref{eq:sigma}) is measured with respect to 
the initial circular flow. The surface density $\Sigma$ is calculated as before 
with an CIC  algorithm. Stars are stable against self-gravitating 
local modes of fragmentation when $Q_J \gtabout 1.7$ (see e.g. Binney \& Tremaine 1987). We applied a modified criterion for stability, because the 
disc is presumed stable against such modes initially, that is, when the black hole is fixed at the 
centre of coordinates and the system is axially symmetric. 
We subtracted from the local mean square velocity 
dispersion the value computed for the symmetric configuration. In this way we 
measure only the increase in dispersion due to black hole motion, and set 
a conservative threshold for stability such that   $Q_J > 1$. When that condition is 
satisfied, the black hole contributes more than 58\% of the square 
velocity dispersion required to prevent local self-gravitating 
fragmentation modes from growing through its orbital motion alone. Since the black 
hole already contributes more than 50\% to the gravity everywhere inside its 
radius of influence, it also provides  the extra dispersion required to kill off all self-gravitating modes. 

  Fig.~\ref{fig:jeans} maps out $Q_J$ at three different times for the reference calculation C3; 
  the shaded area is for  $Q_J > 1$ with an upper cutoff at 2, so white means instability on
  that figure. 
  These images should be compared to their counterpart on Fig.~\ref{fig:density}. 
  It is notable that most of the over-dense thin structures on that figure appear {\it un}stable against 
  fragmentation on Fig.~\ref{fig:jeans} (since they disappear in a sea of white). 
  The outer dark ring at $r \simeq 1.3 $ on the figure matches the position 
  of the 2:5 resonance shown on Fig.~\ref{fig:sos100k} (using $R_c \simeq 0.46$ to revert to 
  physical dimensions). Thus it is very likely that structures that would cross this area would be 
  heated up and disrupted as a result of black hole motion. This may have consequences 
   for the streams of stars observed at the centre of the MW 
   (Genzel et al. 2003). The dimensions
   of this ring, of some $3 r_{bh}$, would correspond to a radius of (roughly) 3 pc for 
   the MW. This should be an element to incorporate into future modelling 
   of the MW centre since actual resolution power already resolves sub-parsec scales. 
    
 \section{Conclusions and future work}
The response of stellar orbits is in direct proportion to the amplitude of motion of a massive 
black hole. The imprint of black hole motion on the stellar kinematics is in direct relation to the 
stars' angular momentum distribution function. Stars on low-angular momentum orbit 
likely will collide with the hole, while those of large momenta experience
strong beat-frequency resonances (when the hole's orbit is either radial or circular). 
 The combined effect left the velocity field significantly anisotropic with a ratio of averaged 
 one-dimensional velocity to rms dispersion reaching $\sim 18\%$. Because analysis suggests that the 
 black hole energy is preserved while that of the stars varies in time, we say the the black hole is a catalyst 
 for evolution of the stellar energy d.f..

The two-dimensional  modelling done in this paper is a first attempt at isolating the generic features of a time-evolving 
dense nuclei with black hole motion. The quantitative outcome of the calculations would be improved in a study 
of a family of anisotropic distribution functions, such as e.g. the Osipkov-Merritt d.f. $f(E-L^2/r_a^2)$ (Binney \& Tremaine 1987, \S4.4.4), using 
self-consistent three-dimensional integrators. We have shown that when stars are 
on circular or near-circular orbits, the resonances induced by the black hole likely will lead to 
self-gravitating substructures inside a volume of a few times $r_{bh}$ in diameter. Such a study carried out with an 
N-body technique is possible provided that collisional physics around the black hole is well resolved (Preto et al. 2004). 
Merritt (2005) and Merritt et al. (2007) have shown that repeated collisions with stars inside $\sim r_{bh}/2$ lead to 
random walk and an effective `Brownian' velocity transferred to the black hole. If the random walk was of amplitude 
$r_{bh}/2$, this would equally imprint the kinetic motion of stars outside $\sim r_{bh}$, as we have seen, through 
the catalytic process that we have outline.   

We have neglected the orbital evolution of the black hole. In reality the stars inside the black hole's radius of influence 
$r_{bh}$ would take away energy and lead to it sinking to the centre through dynamical friction. This does not 
invalidate the impact of black hole motion on the stellar kinematics because 1) this signature is manifest well outside 
$\sim 2 r_{bh}$ and 2) dynamical friction will be effective on a time-scale of $\sim $ few orbital revolutions. 
Recall that the effect discussed here is effective on a single black hole period. 

Black hole orbital 
evolution would bring a higher degree of realism and a more fiducial comparison to observational data. We examined
the case of a black hole on a circular orbit which gave rise to 
an $m = 0$ density wave, spanning a mass of $\sim 40\%$ 
the mass of the black hole. The gravitational torque of the 
 wave would rapidly brake the black hole, which would sink toward the centre and 
lock many stars along with it. If the wave were unstable to collapsing on itself and form a bound object, a double 
nucleus would form. The separation between the two nuclei would be $\sim 2 r_{bh}$ or larger, as deduced from 
the density map on Fig.~\ref{fig:sequencecwh}, panels at the bottom. 
On the contrary, if the tidal field of the BH were too strong, the wave would merge with the 
black hole. This would leave the black hole near the galactic centre surrounded by a pool of 
stars on eccentric orbits. Tremaine (1995) has argued that the 
 double nucleus of M31 (Lauer et al. 1993) 
may be such a case of an off-centre supermassive BH surrounded by a stretched Keplerian disc of size 
$\sim r_{bh}/2$ (0.5'' separation at 800 pc, with $M_{bh} \simeq 8\times 10^7 M_\odot$). Our calculations  did 
not include self-gravity, and hence the fate of the $m = 0$ density wave seen on Fig.~\ref{fig:sequencecwh} remains undetermined. (See also Peiris \& Tremaine 2003, Salow \& Statler 2001, and Bender et al. 2005
for further data on M31.) Other double or multiple nuclei detected in external galaxies 
(e.g., NGC4486B and NGC4382, Lauer et al. 1996, 2005; the Virgo Cluster dwarf VCC 128, Debattista et al. 2006) are prime 
examples of the strong orbital coupling of stars with a supermassive black hole and its influence 
 on the small-scale morphology of a galaxy. New data may reveal cases where double-nuclei 
 galaxies result from the orbital coupling we have discussed here. For the Milky Way, current and future 
 high-precision astrometric missions, such as RAVE 
 (accuracy of $\sim 1$ km/s out to 8 kpc at a magnitude limit of $\sim 13$ [{\it I}-Band] ) or GAIA  (launch date 2011) 
 should pick up any systematic trends in stellar kinematics and set firm constraints on any black hole motion inferred  from stellar streams.  
 
\section*{Acknowledgments}
 We would like to thank T. Lauer and V. Debattista for comments on an earlier version of this 
 paper. CMB was awarded a travel grant from the Indo-French Astronomy Network 
 which made possible a visit to IUCAA in January 2007. 
 Our warm thanks to IFAN's A.K. Kembhavi and A. Lecavelier des Etangs for support.

% \section*{References}

\end{document}